\documentstyle[12pt]{article}
\def\lacute{\mathopen{<}}
\def\racute{\mathclose{>}}
\title{{\bf Chiral spinors and gauge fields in noncommutative curved space-time}}
\author{{\bf Nguyen Ai Viet} \footnote {\noindent E-mail:nguyenaiviet@iop.ncst.ac.vn}\\
Department of High-Energy Physics, CTP, Institute of Physics,\\
National Centre of Science and Technology, Hanoi, Vietnam \\
{\bf Kameshwar C. Wali} \footnote {\noindent E-mail:wali@physics.syr.edu} \\
         Physics Department, Syracuse University \\
         Syracuse, NY  13244-1130, U.S.A. }

\begin{document}
\maketitle
\begin{abstract}
The fundamental concepts of Riemannian geometry, such as
differential forms, vielbein, metric, connection, torsion and
curvature, are generalized in the context of non-commutative
geometry. This allows us to construct the Einstein-Hilbert-Cartan
terms, in addition to the bosonic and fermionic ones in the
Lagrangian of an action functional on non-commutative spaces. As
an example, and also as a prelude to the Standard Model that
includes gravitational interactions, we present a model of chiral
spinor fields on a curved two-sheeted space-time with two distinct
abelian gauge fields. In this model, the full spectrum of the
generalized metric consists of pairs of tensor, vector and scalar
fields. They are coupled to the chiral fermions and the gauge
fields leading to possible parity violation effects triggered by
gravity.
\end{abstract}

\section{Introduction}
    It is widely recognized that our present concepts of space and time are inadequate to
provide a satisfactory basis for a unified description of all
elementary particle interactions including gravity. The twin
pillars of modern physics, namely, 1). General Relativity, in
which the dynamics of the classical continuum of space-time is
coupled to the dynamics of the matter moving within it and 2).
Quantum Field Theory, with rules of quantization to be applied, in
principle, to all degrees of freedom including gravity, are found
to be incompatible. While supersymmetric superstring theory has
provided new insights, it is as yet far from being a convincing
physical theory with predictable and experimentally verifiable
consequences.

    Both general relativity and quantum field theories assume that space-time
is a continuum. A pseudo-Riemannian manifold, based on such a
continuum picture, provides the basis for the geometrical
description of the general theory of relativity. Likewise, quantum
fields and their interactions are local operators that are
functions of continuous, space-time coordinates. If one wants to
explore the possibility of a unified quantum theory of space, time
and matter, it appears that such a continuous space-time is
inadequate. Recent ideas based both on string theory and
approaches to quantum gravity suggest strongly that space-time
structure at the Planck scale may be discrete with noncommutative
coordinates. The continuum space-time of classical physics, it is
hoped, will emerge in certain limiting regimes, just as the
classical behavior of quantum systems emerges in an appropriate
limit.

    At the present time we do not have a precisely defined
noncommutative space that meets the above requirements. However,
in recent years, Connes has proposed an alternate approach to the
study of the structure of space-time, based on noncommutative
geometry (NCG) \cite{c:ncg,c:ess}\footnote { see also \cite{
Coquereaux, gianni, madorebook} for a review}. It has given rise
to the description of the Standard Model with a geometrical
interpretation of the Higgs field on the same footing as the gauge
field \cite{colot,dkm}. Spontaneous symmetry breaking follows as a
natural consequence. It enables one, in principle, to calculate
some arbitrary parameters of the Standard Model, such as the
Weinberg angle, the top quark and Higgs masses \cite{vgb, ZZbar,
WIGNER}.

    Connes' mathematical framework is general enough to explore also the character of
noncommutative spaces at the energy scales that go beyond the
electro-weak scale. In this framework, Connes bypasses the precise
specification of the manifold as the starting point. Instead, he
formulates its description in terms of an associative and
involutive algebra, commutative or noncommutative. One may think
of this as a generalization of the well-known theorem due to
Gelfand \cite{c:ncg, s}, which states that the classical
topological space based on a continuum can be completely recovered
from the abelian algebra of smooth functions. We shall discuss
Connes' ideas in some detail later in this paper, but we note here
that his noncommutative geometry begins with what is called a
spectral triple consisting of i) an involutive algebra ${\cal A}$
of operators, commutative or noncommutative, ii) a Hilbert space
${\cal H}$ as the carrier space for a representation of
generalized differential forms constructed from this algebra, and
iii) a self-adjoint operator $D$, called the Dirac operator,
acting on ${\cal A}$. The algebra, ${\cal A}$, replaces and
generalizes the commutative algebra of smooth functions. The Dirac
operator $D$ allows one to build a differential structure
associated with any associative algebra and defines the metric.

    Although the action functionals, in principle, can be constructed using the metric
structure encoded in the Dirac operator, the procedure is still
not unique. It allows different approaches to different problems.
Thus, while the Diximier trace is used successfully to construct
the action functional for the gauge sector of the Standard Model,
other approaches have been pursued in the literature in dealing
with the gravitational interactions. The Wodzicki residue,for
instance, is proposed to be used as an action functional for the
gravity sector in the two-sheeted space-time. It leads to the
conventional Einstein-Hilbert action supplemented by a term
proportional to the square of a scalar field \cite{wodzicki,
kastler}. The vector field that is part of the generalized metric
is cancelled completely in the final action. The scalar field
without a kinetic term does not propagate and causes problems when
coupled to matter fields. The Diximier trace is used in other
approaches with different generalizations of the Cartan formalism
leading to a Brans-Dicke field coupled to gravity
\cite{chff,sitarz}. In these approaches, the vector field vanishes
as a consequence of a reality condition or a consistency
condition. In a more recent paper \cite{cch}, Connes and
Chamseddine have proposed a new universal spectral action
principle governing a noncommutative space that gives rise to the
Einstein-Hilbert action together with higher order terms.

    Landi, Viet and Wali \cite{LVW} have adopted still another approach. It follows closely
the standard Riemanian geometry and generalizes it in the case of
a two-sheeted space-time that represents a discretized version of
Kaluza-Klein theory in which, the fifth compact circular dimension
is replaced by two discrete points. By formulating the basic
notions of the conventional geometry in terms of algebraic forms,
they define the latter in such a way that they make sense both in
the case of commutative and noncommutative situations. In their
simplified version of the model in which the generalized metric is
assumed to be the same on the two sheets,the resulting action
functional is exactly that of the zero mode sector of the
Kaluza-Klein theory. In the more general case studied by Viet and
Wali \cite{VW2, VW1}, the Cartan formalism of pseudo-Riemannian
geometry finds its generalization to allow the most general metric
that contains pairs of tensor, vector and scalar fields. All of
these fields are present in the resulting action functional with a
set of constraints on the non-vanishing torsion that generalizes
the torsion free condition in the standard space-time. They find a
finite spectrum of the discretized Kaluza-Klein theory without
requiring the truncation of the infinite mass spectrum. In the
special case \cite{VW1}, when the torsion free condition is
imposed, the spectrum reduces to the zero mode sector coupled to
dilatons with a sixth order potential and a cosmological constant
\cite{VW1}. Recently, Viet \cite{Vtorsion} has found a new minimal
set of constraints, which together with the Cartan equations
determine the connection, torsion and curvature in terms of the
pairs of tensor, vector and scalar fields with a simpler
Einstein-Hilbert-Cartan action functional.

In this paper, we extend our formalism to include a fermionic and
its associated gauge boson sectors. As a prelude to the study of
the Standard Model to include gravity, we consider a model of a
left- and a right- chiral field living on a curved two-sheeted
space-time and construct their action functionals. Due to the
existence of different metric fields on the two sheets of
space-time, we find a possible mechanism in our formalism for
parity violation due to gravity.

In Section 2, we present the basic elements of the noncommutative
differential geometry after a review of the algebraic approach to
the standard differential geometry.  This is primarily to set the
stage for the contrasting NCG approach. The formulas of the key
geometric notions in the standard case are formulated in such a
way that they can be used as guidelines for generalizations in the
noncommutative situation. In the Section 3., continuing the same
approach, we introduce metric structure and construct Riemannian
geometry using the algebraic approach and generalize it to the
case of noncommutative geometry \`a la Connes. We also discuss in
this section the representation mapping $\pi$ given by Connes
\cite{c:ncg} to represent operators in a given Hilbert space. This
procedure will be important in the realization of a noncommutative
geometric model in the rest of the paper. In Section 4., we
specialize to the algebra of differential forms and metric
structure on the two-sheeted space-time. Here we follow the
minimal constraints given in \cite{ Vtorsion} to construct the
Einstein-Hilbert-Cartan action. By following closely, the formulas
given in Section 3, we establish a concrete realization of the
abstract noncommutative geometry \`a la Connes in our formalism.
In Section 5., we discuss the physical contents of the gravity
sector in two-sheeted space-time. In Section 6., we construct the
action for the gauge and chiral fermion sectors. Section 7. is
devoted to discussions of some physical consequences of the model,
and the final section is devoted to a summary and conclusions.

\section{ Algebraic approach to differential geometry}

\subsection{ Global and algebraic approaches versus local construction}
\hskip 0.6cm
     The traditional starting point of the standard differential geometry is the local construction of the vector spaces
of tangent vectors and differential forms in the coordinate bases
at an arbitrary local point $x \in {\cal M}$
\cite{chandra,misner,wald,nakahara}. In this construction, the
differential forms are formulated as duals to the tangent vectors,
which are derivatives of the functions at a point $x$. The
differential forms form a vector space. All the geometric notions
such as  metric, connection, torsion, curvature and physical
fields as scalars, vectors, tensors can be formulated conveniently
in terms of differential forms. Although these notions can be
defined locally, their meaning can only be understood clearly in
the context of a global construction \cite{nakahara}.

    The physical fields that are defined at all points of a manifold ${\cal M}$, should be formulated strictly as
the sections in a global framework of fiber bundles. In
particular, the spinors are meaningful if and only if a global
property of the manifold ${\cal M}$, the orientability, can be
defined. The physical fields, which are defined as sections of the
fiber bundles with ${\cal M}$ as the base manifold, have an
algebraic finite projective module structure. The connection
together with other concepts such as the covariant derivative and
the curvature can also be defined in the framework of this global
construction \cite{nakahara}. This observation is important as it
paves the way to a pure algebraic approach that allows direct
generalization of the ordinary geometry to NCG without reference
to the concept of a point.

    The concept of a point as the starting point is the main obstacle to go beyond
the standard geometry and seek a unified framework of general
relativity and quantum theory. In an algebraic approach, on the
other hand, it is possible to start from the pure algebraic
structures to formulate the basic geometric elements. Thus, the
Gelfand's theorem \cite{s}, which so far has been just an
alternative formulation of differential geometry, becomes the
central point in such an algebraic approach to geometry without
the concept of a point.

    A discussion of the global approach could provide a great deal of useful
insight because it can serve as a bridge between the traditional
and algebraic constructions. However, such a task is beyond the
scope of this paper. \footnote{The interested readers are referred
to \cite{nakahara} for more details in physical applications.}
However, from the algebraic relations that are originally derived
from a local construction of the ordinary geometry, the basic
geometric and physical elements can be formulated in such a way
that they can be generalized directly to NCG. In the following
Sections, it is our goal to reformulate the basic notions of the
ordinary geometry in the algebraic approach first, and then
directly generalize them  to the non-commutative case.

\subsection{ The ordinary differential geometry}
\hskip 0.6cm
    The essential point of an algebraic approach is taking the algebra of function
    as a starting point. The set of differential forms will be considered as an algebraic
module, a concept more general than that of a vector space, which
also makes sense when the algebra is noncommutative.

    The continuous functions $f(x):{\cal M} \longrightarrow R$ form a commutative algebra $C^\infty({\cal
    M})$.According to
 Gelfand's theorem, all the information about the manifold ${\cal
M}$ is encoded in this algebra. In other words, knowing the
algebra $C^\infty({\cal M})$ is equivalent to knowing the manifold
${\cal M}$ in the ordinary differential geometry.

    The graded algebra of differential forms is then constructed with the help of a differential
    operator $d$ that maps the
function $f$ as a 0-form to the 1-form $df$. The module of 1-forms
is extended by including all the elements $df.g$, where $g$ is an
arbitrary element of $C^\infty({\cal M})$. The higher forms can be
defined by extending the differential operator $d$ as the mapping
from $n$-forms onto $n+1$-forms.

        The operator $d$ is required to satisfy the following properties: i) $d^2 = 0$,
ii) $d(ab) = da.b + (-1)^n a.db$, where $a$ is a $n$-form and $b$
is an arbitrary form (Leibnitz rule), iii) $d$ is hermitian.

        In general, the differential operator can be given as any operator that satisfies
        the above three properties. In
the particular case of ordinary differential geometry, the
differential operator $d$ is given in a finite basis of the
1-forms $dx^\mu$ ($\mu = 0,1,...,n$), \setcounter{equation}{1}
$$
d = dx^\mu \partial_\mu. \eqno{(\thesection .\theequation)}
\label{2.1}
$$
      The assumption about the existence of a finite basis of a module of 1-forms is essential in the
non-commutative case.

    The action of $d$ on a function $f$ as a c-number operator is given as follows
\refstepcounter{equation}
$$
df = [d,f] = dx^\mu \partial_\mu f(x). \eqno{(\thesection
.\theequation)} \label{2.2}
$$

An arbitrary 1-form $u$ is also generated by  the elements $dx^\mu
$ \refstepcounter{equation}
$$
u = dx^\mu u_{\mu}, \eqno{(\thesection .\theequation)} \label{2.3}
$$
where $u_\mu$  are contained in $C^\infty({\cal M})$.

    The action of $d$ on $u$ in accordance with the properties i)-iii) is
\refstepcounter{equation}
$$
du = d (dx^{\mu} u_{\mu}) = d(dx^{\mu}). u_{\mu} - dx^{\mu}
dx^{\nu} \partial_{\nu} u_{\mu} = - dx^{\mu} dx^{\nu}
\partial_{\nu} u_{\mu}. \eqno{(\thesection .\theequation)} \label{2.4}
$$

We can write $du$ as a combination of an antisymmetric and a
symmetric 2-forms, \refstepcounter{equation}
$$
du = {1\over 4}(dx^\mu dx^\nu - dx^\nu dx^\mu) (\partial_\mu u_\nu
- \partial_\nu u_\mu) - {1\over 4}(dx^\mu dx^\nu + dx^\nu
dx^\mu)(\partial_\mu u_\nu + \partial_\nu u_\mu).
\eqno{(\thesection .\theequation)} \label{2.5}
$$

   The symmetric 1-forms must be dropped since they lead to the so
   called``junk forms''. In a specific representation, a non-zero form can
be derived from a form which is identical to zero. Junk forms are
such forms. As an example, consider the form $df.f - f.df$ that is
identical to zero; however its differential $d ( df.f - f.df) =
-df.df - df.df = -2df.df $ is not zero, if the symmetric form is
allowed. The algebra of anti-symmetric 1-forms is a closed algebra
by itself.

  There are various ways to avoid junk forms. The simplest way, however, is to
  replace
 an ordinary product by the wedge product

\refstepcounter{equation}
$$
dx^\mu \wedge dx^\nu = {1\over 2} (dx^\mu dx^\nu - dx^\nu dx^\mu).
\eqno{(\thesection .\theequation)} \label{wedge}
$$

    The Eqn.(2.\ref{wedge}) can be generalized directly for the wedge product of an arbitrary
number of the elements $dx^\mu$ just by anti-symmetrization.

    A n-form $\omega$ is a purely antisymmetric tensor of the following form
\refstepcounter{equation}
$$
\omega = dx^{\mu_1} \wedge ...\wedge dx^{\mu_n}
\omega_{\mu_1...\mu_n}. \eqno{(\thesection .\theequation)}
\label{2.7}
$$

    The $n$-forms form a vector space $\Omega_x^n({\cal M})$. The wedge product is defined between a $m$-form $\omega_1$
and a $n$-form $\omega_2$ to form a $m+n$-form $\omega_1 \wedge
\omega_2$ as an anti-symmetrized tensor product

\refstepcounter{equation}
$$
\wedge : \Omega_x^n({\cal M}) \times \Omega_x^m({\cal M})
\longrightarrow \Omega_x^{m+n}({\cal M}). \eqno{(\thesection
.\theequation)} \label{2.8}
$$

    One can define the graded algebra of the differential forms as
$\Omega_x({\cal M}) = \bigoplus_n \Omega^n_x({\cal M})$.

    As an example of differential forms, a Lie-algebra valued gauge field belonging to a group
    $G$ or gauge connection $b$ ,can be given as the 1-form
\refstepcounter{equation}
$$
b = dx^\mu b_\mu , \eqno{(\thesection .\theequation)} \label{2.9}
$$
with its field strength as the 2-form \refstepcounter{equation}
$$
f = db + b \wedge b = dx^\mu \wedge dx^\nu  f_{\mu \nu} = dx^\mu
\wedge dx^\nu {1\over 2} [(\partial_\mu b_\nu - \partial_\nu
b_\mu) + [b_\mu, b_\nu] ]. \eqno{(\thesection .\theequation)}
\label{2.10}
$$

     In a "semi-classical" limit, all the forms are defined as
    operators on a Hilbert space
${\cal H}$ of spinor wave functions. The strict geometric
formulation of the physical spinor fields is given as spinor
bundles. On the Hilbert space ${\cal H}$, we can construct a
$\gamma$- matrix representation by identifying the abstract
element $dx^\mu$ as \refstepcounter{equation}
$$
dx^\mu \equiv \gamma^\mu, \eqno{(\thesection .\theequation)}
\label{2.11}
$$
where $\gamma^\mu$ are the usual Dirac matrices.

In the $\gamma$-matrix representation, the differential operator
$d$ is represented by the Dirac operator \refstepcounter{equation}
$$
d = \gamma^{\mu} \partial_{\mu} . \eqno{(\thesection
.\theequation)} \label{2.12}
$$

The gauge field one-form and the field strength two-form are
represented respectively by
\refstepcounter{equation}
$$
b = \gamma^{\mu} b_{\mu}, \eqno{(\thesection .\theequation)}
\label{2.13}
$$

and
\refstepcounter{equation}
$$
f = \sigma^{\mu \nu} f_{\mu \nu}, \eqno{(\thesection
.\theequation)} \label{2.14}
$$
where $\sigma^{\mu \nu} = {1 \over 2} [\gamma^\mu, \gamma^\nu]$.

  Riemannian geometry can be constructed, in accordance with the Equivalence Principle,
   by introducing a locally flat
orthonormal basis $\gamma^a, a=0,1,2,3 $ and introducing a
vierbein structure, which gives the transformation between the two
frames \refstepcounter{equation}

$$
\begin{array}{ccc}
 \gamma^\mu & = & e^\mu_a \gamma^a \\
 \gamma^a   & = & e^a_\mu \gamma^\mu \\
 e^a_\mu e^\mu_b & = & \delta^a_b \\
 e^a_\mu e^\nu_a & = & \delta^\mu_\nu
 \end{array}
\eqno{(\thesection.\theequation)} \label{2.14aa}
$$

   Gravity can be introduced via the connection 1-forms
\refstepcounter{equation}
$$
\omega_{ab} = \gamma^c \omega_{abc}.
\eqno{(\thesection.\theequation)} \label{2.14a}
$$

   Two main notions that characterize the gravity structure in space-time,
   the torsion and curvature, can be defined by the Cartan's structure equations.

The torsion 1-forms $t^a$ are given in the first structure
equation

\refstepcounter{equation}
$$
t^a  = d\gamma^a - \gamma^b \wedge \omega^a_b
\eqno{(\thesection.\theequation)} \label{2.14b}
$$

The curvature 2-forms $ r^a_{~b} $ given by the second structure
equation \refstepcounter{equation}
$$
r^a_{~b}  = d\omega_{ab} + \omega^a_{~c} \wedge \omega^c_{~b},
\eqno{(\thesection.\theequation)} \label{2.14c}
$$
determines the scalar curvature completely in terms of
connections.

\subsection{ Non-commutative differential geometry and the spectral triple}

\hskip 0.6cm
  In the case of  non-commutative geometry, we shall follow the same
steps as outlined in Sect 2.2. We will begin with Connes' spectral
triple \cite{c:ncg,c:ess}, which consists of :
   1) An involutive, unital algebra of functions ${\cal A}$,
   2) Hilbert space ${\cal H}$, and
   3) Dirac operator $\delta$ that satisfies the properties
   i)-iii)in Sect.2.2.

   The existence of a finite basis for the module of differential forms is
understood. Also, the algebra ${\cal A}$ of generalized functions
is not necessarily commutative as in the conventional case. In
this subsection, we shall begin with a brief review of Connes'
abstract construction of universal enveloping algebra of the
noncommutative differential forms.

    The universal differential enveloping algebra constructed from a noncommutative
algebra ${\cal A}$ is the differential algebra $\Omega^*({\cal
A})$, generated by all $a, b \in {\cal A}$ and a symbol $\delta$,
such that \refstepcounter{equation}
$$
   \delta(1) = 0 ,
$$
$$
   \delta(ab) = (\delta a)b + a(\delta b).
\eqno{(\thesection .\theequation)} \label{Leibnitz}
$$

    By definition, the algebra of 0-forms
    $\Omega^0({\cal A}) \equiv {\cal A}$; $\delta a$
belongs to the space of universal 1-forms $\Omega^1({\cal A})$,
whose general element $\omega$ is a linear combination

\refstepcounter{equation}
$$
\omega = \sum_i \delta a_i b_i , \hspace{.5cm} a_i, b_i \in {\cal
A}. \eqno{(\thesection .\theequation)} \label{Gen-1form}
$$

    The existence of a finite number of elements $\delta a_i$ is guaranteed
by a generalization of the vector space to the finite projective
module. More details about this concept can be found in
\cite{c:ncg, Coquereaux}. However, here we just remark that in the
ordinary geometry there is also a finite number of elements
$dx^\mu$ which are the particular forms of $\delta a_i$.

    In a vector space, scalar multiplication of a vector is defined with elements
of an algebraic field, usually that of the complex or real
numbers. The direct generalization of a vector space of 1-forms is
an ${\cal A}$-module, where the linear combinations with real or
complex coefficients are replaced by the elements of ${\cal A}$
and the scalar multiplication is defined with elements of the
algebra ${\cal A}$ from right

\refstepcounter{equation}
$$
\omega b = (\sum_i \delta a_i b_i) b =\sum_i \delta a_i  (b_i
b),\hspace{.5cm} \forall b \in {\cal A}. \eqno{(\thesection
.\theequation)} \label{2.17}
$$

    The Leibnitzian property of $\delta$, Eqn.(2.\ref{Leibnitz}), can be
used to express left multiplication of $\omega$ by an element
$b\in {\cal A}$,

\refstepcounter{equation}
$$
\begin{array}{ccl}
b \omega  & = & b (\sum_i \delta a_i  b_i) \\
& = & \sum_i \delta (b a_i ) b_i - \delta b a_i b_i ,\hspace{.5cm}
\forall a_i, b_i, b \in {\cal A}.
\end{array}
\eqno{(\thesection .\theequation)} \label{leftmodule}
$$

    That is to say, the generalized 1-forms give rise to an ${\cal A}$-bimodule structure.
Here, for the sake of definiteness, we have used the basis with
$\delta a_i$ on the left. However, it is always possible to use
the Leibnitz rule (2.\ref{Leibnitz}) to rewrite a differential
form with the right basis. Due to the non-commutativity of the
algebra ${\cal A}$, the basis elements and the expansion
coefficients in the right basis are not necessarily the same as in
the left one as in the commutative case.

To proceed further,we note that we can multiply two 1-forms as in
the commutative case. The associativity of the algebra ${\cal A}$
and Eqn.(2.\ref{Leibnitz}) imply \refstepcounter{equation}
$$
(\delta b_0 a_0) (\delta b_1 a_1)= \delta b_0 (a_0 \delta b_1) a_1
=\delta b_0 \delta (a_0 b_1) a_1 -\delta b_0 \delta a_0 (b_1 a_1).
\eqno{(\thesection .\theequation)} \label{2.19}
$$

   Continuing in this manner, we can write p-fold products of 1-forms,
 ${\cal A}$-coefficient linear combinations of which give the algebra $\Omega^p({\cal A})$.
A general element of $\Omega^p{\cal A}$ is of the form
\refstepcounter{equation}
$$
\sum_i \delta a_{1i} ... \delta a_{pi} b_i ,\hspace{.5cm}a_{ki},
b_i \in {\cal A}. \eqno{(\thesection .\theequation)} \label{2.20}
$$
    The algebra $\Omega^p({\cal A})$ is now a right ${\cal A}$-module being
    a
generalization of the vector space of p-forms on a manifold.
Clearly, by applying repeatedly Eqn.(2.\ref{leftmodule}), one may
define the product of a universal p-form with a q-form
\refstepcounter{equation}
$$
\Omega^p{\cal A} \times \Omega^q{\cal A} \mapsto \Omega^{p+q}{\cal
A}  . \eqno{(\thesection .\theequation)} \label{2.21}
$$

    Hence, the space of universal forms has the structure of a graded algebra, the
universal algebra $\Omega^*({\cal A}) \equiv \bigoplus_p
\Omega^p{\cal A}$. The involution on ${\cal A}$ extends uniquely
to an involution on the algebra $\Omega^*{\cal A}$ when we impose
the condition \refstepcounter{equation}
$$
   (\delta a)^* \equiv - \delta a^*
\eqno{(\thesection .\theequation)} \label{2.22}
$$

    To transform the graded algebra of forms $\Omega^*{\cal A}$ into a differential
algebra, we consider the ${\it differential}$ $\delta $ as a
linear operator that takes \refstepcounter{equation}
$$
\Omega^p \longrightarrow \Omega^{p+1} \eqno{(\thesection
.\theequation)} \label{2.23}
$$
by defining \refstepcounter{equation}
$$
\delta (\delta a_1 ... \delta a_p b) \equiv  (-1)^p \delta a_1 ...
\delta a_p \delta b. \eqno{(\thesection .\theequation)}
\label{Dact-pform}
$$

    Eqn.(2.\ref{Dact-pform}) implies two basic relations,
\refstepcounter{equation}
$$
\delta^2 \alpha = 0 , \hspace{.5cm}\forall \alpha \in
\Omega^*{\cal A}, \eqno({\thesection .\theequation)} \label{2.30}
$$
\refstepcounter{equation}
$$
  \delta (\alpha_1 \alpha_2) = (\delta \alpha_1)\alpha_2 +
(-1)^{deg \alpha_1}\alpha_1\delta \alpha_2 ,\hspace{.5cm} \forall
\alpha_j \in \Omega^*{\cal A}. \eqno({\thesection .\theequation)}
\label{Leibnitz-pform}
$$
    Thus, all the constructions of the ${\cal A}$-module of differential
forms are in exact parallel with the standard approach. In the
next section, we shall present the general procedure to realize
the above abstract construction in a given Hilbert space. We note
that in the module of the generalized differential forms, the
``junk'' forms must be eliminated to guarantee that if $\delta
\omega \not= 0 $ the differential form $\omega $ must be a
non-zero form.

\subsection{ Representation of the differential algebra on a given Hilbert space}
\hskip 0.6cm
  The general constructions described in Sect 2.3 need to be realized by operators acting on
appropriate Hilbert spaces in different applications. The operator
representation of the differential forms is a direct
generalization of the $\gamma$- matrix representation given in
Sect. 2.2.

 Connes \cite{c:ncg, c:ess, Coquereaux} has given a general procedure
to construct representations by the following graded homomorphism
that preserves the involution,
$$
\pi : \Omega^*{\cal A} \longrightarrow {\cal L}({\cal H}),
$$
\refstepcounter{equation}
$$
\pi_p(\delta a_1 ...\delta a_p b)= \prod_{i=1}^p[D,\pi _0 (a_i)]
\pi _0(b), \eqno{(\thesection .\theequation)} \label{piref}
$$
where ${\cal L}({\cal H})$ denotes the space of bounded operators
on ${\cal H}$ and $\pi_0$, the representation of ${\cal A}$ on
${\cal H}$. The operator $\delta$ is represented by a self-adjoint
operator $D$, the Dirac operator, on the Hilbert space ${\cal H}$,
with compact resolvent, such that the commutator $[D,a]$ is a
bounded operator, $\forall a \in {\cal A}$.

In essence, differential forms, in the sense of Connes, are the
images of universal forms under $\pi$. Strictly, however, a
problem arises in the definition of these operator-valued forms,
in that there exist universal forms $\alpha$ such that $\pi
(\alpha )=0$, but $\pi (\delta \alpha) \not=0$. Once one mods out
by these troublesome forms for each p, we obtain the space of
operator-valued forms $\Omega^p_D({\cal A})$, the analog of
$\Omega^p({\cal A})$. This is the problem of elimination of the
``junk'' forms in ordinary geometry. In two-sheeted space-time, as
it will be shown in Sect. 4, this elimination can be done by a
generalized wedge product in perfect parallelism with the
conventional approach.

The various definitions proceed as in the previous subsection. We
define

\refstepcounter{equation}
$$
\Omega^*_D({\cal A}) \equiv \bigoplus \Omega^p_D({\cal A})
\eqno{(\thesection .\theequation)} \label{2.28}
$$

Consider Eqn.(2.\ref{Gen-1form})as a specific case to illustrate
the representation given in Eqn.(2.\ref{piref}). The arbitrary
1-form $\omega$ can be represented as an operator $U$ on the
Hilbert space ${\cal H}$ as follows \refstepcounter{equation}
$$
U = \pi_1(\omega) = \pi_1(\delta a_i) \pi_0(b_i) = \Theta^i U_i,
\eqno{(\thesection .\theequation)} \label{2.29}
$$
where
 $\Theta^i$ are just the representations of
the element $\delta a_i$ in Eqn.(2.\ref{Gen-1form}). Thus,in this
general context, the basis of differential forms is given without
reference to a point. To be more specific with the Dirac operator
$D$, let $F$ be an operator representing a generalized function
$w$ in $\Omega^0_D({\cal A}) \equiv {\cal A}$. From
Eqn.(2.\ref{piref}) it follows that

\refstepcounter{equation}
$$
\begin{array}{ccc}
DF  & =  & [D, \pi_0(w)] = \pi_1(\delta w) = \pi_1(\delta a_i w_i) \\
    & =  & \pi_1(\delta a_i) \pi_0(w_i) = \Theta^i (DF)_i.
\end{array}
 \eqno{(\thesection .\theequation)} \label{2.30a}
$$

The coefficients $(DF)_i \in {\cal A}$ can now be regarded as
derivatives but not referring to a specific point. Formally, we
can define the operator $D_i$ as \refstepcounter{equation}
$$
(DF)_i \equiv D_i F = [D_i, F]. \eqno{(\thesection .\theequation)}
\label{2.31}
$$

Hence, we can represent the Dirac operator in the $\Theta^i$ basis
in the form \refstepcounter{equation}
$$
D = \Theta^M D_M . \eqno{(\thesection .\theequation)} \label{2.32}
$$

    From now on we use the latin upper case indices $M,N = 0,...,n-1$
to replace the latin lower case indices $i$. Note that the
expression of the Dirac operator $D$ in terms of the operators
$D_M$, implies a specific basis $\Theta^M$. We are free to choose
another basis of one-forms as a linear combination with
coefficients $\in {\cal A}$. However, once the Dirac operator is
given in a specific basis $\Theta^M$, we can  calculate derivation
of forms in this basis and then transform the result to another
basis whenever necessary.

    Having outlined the construction of the operator algebra $\Omega^*_D({\cal A})$, we
shall henceforth abandon explicit mention of the distinction
between universal forms and their representation as operators on
${\cal H}$, and denote them by capital letters.

\section{Algebraic Formulation of Riemannian geometry
within the framework of NCG}

The construction of the universal algebra of differential forms
based on a pseudo-Riemannian manifold follows the steps discussed
in Sect.2. In its application to the theory of gravitation in
general relativity, we need to introduce the concept of a metric,
the affine connection, the inner products of one and two-forms,
the torsion and curvature. In the NCG approach, the spectral
triple consists of

The algebra ${\cal A}$: The algebra $C^\infty({\cal M}) $ of the
analytic complex functions on the manifold ${\cal M}$.

The Hilbert space ${\cal H} $: Square integral functions that
belong to the direct product of ${\cal A}$ and the spinor bundle
$L^2(S, {\cal M})$.

Dirac operator $ D $, whose representation in the Hilbert space
${\cal H}$ is to be specified.

 The starting point of the general theory of relativity is the
Equivalence Principle, which postulates that in an arbitrary
gravitational field (curved space-time described by Riemannian
geometry),  it is possible to chose a locally inertial system
(flat coordinate system), such that within a sufficiently small
region, the laws of special relativity prevail. The transformation
between the locally flat ortho-normal coordinate system and the
general curved coordinate system introduces the well known
vierbeins that can be regarded as gravitational fields.

The general Equivalence Principle allows the introduction of the
metric via two sets of 1-forms $\Theta^A$ and $\Theta^M$. The
1-forms $\Theta^A$ can be expanded in the $\Theta^M$ basis and
vice versa,

\setcounter{equation}{1}
$$
\Theta^A = \Theta^M E^A_M , \hskip 5mm \Theta^M = \Theta^A E_A^M .
\eqno{(\thesection .\theequation)} \label{EQPRIN}
$$

$E^A_M$ and $E^M_A$ are elements of ${\cal A}$ satisfying
\refstepcounter{equation}
$$
E^A_M E^M_B = \delta^A_B, \hskip 5mm E^A_M E^N_A = \delta^M_N .
\eqno{(\thesection .\theequation)} \label{3.2}
$$

 We define a metric tensor ${\cal G}^{MN}$ as the inner product of two 1-forms $\Theta^M$ and
 $\Theta^N$. The inner product of 1-forms is defined as the sesqui-linear functional

$$
< , > : \Omega^1_D \times \Omega^1_D \longrightarrow {\cal A},
$$

and \refstepcounter{equation}
$$
{\cal G}^{MN} \equiv <\Theta^M, \Theta^N>.
\eqno{(\thesection.\theequation)} \label{INNER}
$$

${\cal G}^{MN}$ are functions that are elements of the algebra
${\cal A}$, the algebra of infinitely smooth functions on the
Riemannian manifold. As the algebra ${\cal A}$ is unital,
$\eta^{AB}$ ( where $\eta^{AB} = diag (-1,1,...,1))$ is also an
element of ${\cal A}$. Therefore, we can choose the one-forms
$\Theta^A$  so that they satisfy \refstepcounter{equation}
$$
{\cal G}^{AB} \equiv < \Theta^A, \Theta^B > = \eta^{AB}.
\eqno{(\thesection .\theequation)} \label{DEFMETRIC}
$$
We reserve Latin uppercase A, B, ... to denote the
``locally-flat'' indices and M, N, ... to denote the ``curved'' or
``derivative'' indices in the basis $\Theta^M$.

   From Eqns.(3.\ref{EQPRIN}) and (3.\ref{DEFMETRIC}) we can derive the relation
\refstepcounter{equation}
$$
\begin{array}{ccc}
 < \Theta^A , \Theta^B > &~=~&\eta^{AB}~=~<\Theta^M E^A_M,
\Theta^N E^B_N >  \\
&~=~& {\tilde E}^A_M <\Theta^M ,\Theta^N > E^B_N ~= ~ {\tilde
E}^A_M {\cal G}^{M N } E^B_N,
\end{array}
\eqno{(\thesection .\theequation)} \label{ORTH}
$$
where, ${\tilde E}^A_M$ denotes the involutive operation if it is
defined for the algebra ${\cal A}$, otherwise it is $E^A_M$.

   We note that the metric structure is determined if the curvilinear
and locally orthonormal bases are given. The vierbein as a
transformation matrix between the two bases determines the metric
tensor. In the algebraic construction, it is more convenient to
use these bases as a starting point to construct the Riemannian
geometry.

If U is a general 1-form, $ U =\Gamma^M U_M = \Gamma^A U_A $, the
transformation between its components in the two coordinate
systems is given by

\refstepcounter{equation}
$$
U_M = E^A_M U_A ~~~,~~~ U_A = E^M_A U_M,
\eqno{(\thesection.\theequation)} \label{3.6}
$$
where $E^M_A$ is the inverse matrix of $E^A_M$.

In particular, in the locally orthonormal basis, the exterior
derivative of arbitrary forms must be first calculated  in the
curvilinear basis and then transformed back to the locally flat
one by using the above transformations. For instance,

\refstepcounter{equation}
$$
(DF)_A= E^M_A D_M F, \eqno{(\thesection .\theequation)}
\label{3.7}
$$

\refstepcounter{equation}
$$
(DU)_{AB}= {1 \over 2}E^M_A E^N_B ( D_M U_N - D_N U_M).
\eqno{(\thesection .\theequation)} \label{3.8}
$$
 The inner product of the basis 1-forms in Eqns.(3.\ref{INNER})and (3.\ref{ORTH})
generates the inner product in the vector space of 1-forms between
$U=\Theta^M U_M$ and $V=\Theta^M V_M$ given by
\refstepcounter{equation}
$$
< U, V >={\tilde U}_M{\cal G}^{MN} V_N = {\tilde U}^N V_N.
\eqno{(\thesection.\theequation)} \label{3.9}
$$

The covariant derivative operator ( affine connection), $\nabla$,
is given by the connection 1-forms $\Omega^A_B$, where
\refstepcounter{equation}
$$
\nabla\Theta^A = \Theta^B\bigotimes \Omega^A_B =
\Theta^B\bigotimes\Theta^C\Omega^A_{BC}.
\eqno{(\thesection.\theequation)} \label{COVDEV}
$$

     The Cartan structure equations define the torsion and curvature 2-forms of a given connection
$\Omega_{AB}$ as follows:

\refstepcounter{equation}
$$
T^A = D\Theta^A - \Theta^B \wedge \Omega^A_B, \eqno{(\thesection
.\theequation)} \label{STRUCT1}
$$
\refstepcounter{equation}
$$
R^A_B = D\Omega^A_B + \Omega^A_C\wedge \Omega^C_B.
\eqno{(\thesection .\theequation)} \label{STRUCT2}
$$

   In Eqns.(3.\ref{STRUCT1}) and (3.\ref{STRUCT2}), we need a definition of the wedge
product of 1-forms. Such a definition, as explained in Sect.2,
becomes specific when the representation mapping $\pi$ is given.
Here we only assume that the module of 2-forms is spanned by the
formal wedge products $\Theta^A\wedge \Theta^B$ of 1-forms. For
the ordinary space-time components, the anti-symmetrization is the
standard convention. However, there is no requirement that it must
be the case for other generalized components. In general, there
are various ways to define the wedge product, depending on the
specific applications.

 The Cartan structure equations (3.\ref{STRUCT1}) and (3.\ref{STRUCT2})
are not sufficient to determine connection, torsion and curvature
in terms of the vierbeins. In the conventional differential
geometry, one generally imposes two additional constraints: i) the
metric compatibility condition

\refstepcounter{equation}
$$
\Omega_{AB} = - \Omega_{BA}
\eqno{(\thesection .\theequation)}
\label{3.13}
$$
and ii) the torsion free condition

\refstepcounter{equation}
$$
T^A = 0 . \eqno{(\thesection .\theequation)} \label{3.14}
$$

 With these two constraints, the first structure equation determines the
connection completely in terms of the vierbeins. The second
structure equation, in turn, determines the curvature in terms of
the connection.

   To calculate the scalar curvature from the curvature two-form, Eqn.(3.\ref{STRUCT2}),
the inner product of 2-forms is introduced as given by
\refstepcounter{equation}
$$
\lacute \Theta^A \wedge \Theta^B, \Theta^C \wedge \Theta^D \racute
\equiv \eta^{AD} \eta^{BC} -\eta^{AC} \eta^{BD}.
\eqno{(\thesection .\theequation)} \label{3.15}
$$
    This inner product can also be extended for two arbitrary 2-forms
$G \equiv \Theta^A \wedge \Theta^B G_{AB}$ and $H \equiv \Theta^A
\wedge \Theta^B H_{AB}$, $G_{AB}, H_{AB} \in {\cal A}$ as follows:
\refstepcounter{equation}
$$
\lacute G, H \racute =  {\tilde G}_{AB}\lacute \Theta^A \wedge
\Theta^B, \Theta^C \wedge \Theta^D \racute H_{CD}.
\eqno{(\thesection .\theequation)} \label{3.16}
$$

Finally, we note that the $\Gamma $-matrix representation is given
by the following homomorphism $\Gamma$

\refstepcounter{equation}
$$
\Gamma(\Theta^A)=\Gamma^A,~~~~~ \Gamma(\Theta^M)=\Gamma^M,
\eqno{(\thesection.\theequation)}\label{GREP}
$$
where $\Gamma^A$ are given as constant "flat" Dirac matrices. The
concrete form of the "curved" Dirac matrices $\Gamma^M(x)$ is not
known since it involves gravitational degrees of freedom through
the vielbeins,

\refstepcounter{equation}
\
$$
\Gamma^M(x)= E^M_A\Gamma^A.
\eqno{(\thesection.\theequation)}\label{3.18}
$$

Any specific assumption about the form of $\Gamma^M$ independent
of the above equation will be an arbitrary restriction on the
metric structure \footnote{ In spite of this obvious fact, this
kind of hidden assumption is often made in the literature without
an explicit statement.}.

With the defining homomorphism (3.\ref{GREP}) , we can transform
the various formula in the $ \Theta $ basis to the corresponding
ones in the $\Gamma$-representation basis. Thus, for instance, the
metric tensor is given by the inner product
\refstepcounter{equation}
$$
G^{MN} =<\Gamma^M , \Gamma^N> = {1 \over 4}Tr(\Gamma^M \Gamma^N),
\eqno{(\thesection .\theequation)} \label{3.19}
$$
where the trace is taken over the spinor indices. From
Eqn.(3.\ref{ORTH}), the expression of the metric tensor in terms
of the vierbeins is given as

\refstepcounter{equation}
$$
{\cal G}^{MN} = {\tilde E}^M_A \eta^{AB} E^N_B. \eqno{(\thesection
.\theequation)} \label{3.20}
$$

Other details will be given in the next section where we use
explicitly the ${\Gamma}$-matrix representation.

\section{ Geometry in two-sheeted space-time}

  In this section, we shall apply the general formalism developed
  in the previous section to the case of a two-sheeted space-time
  with a specific choice for the spectral triple.

\subsection{ Noncommutative differential forms in two-sheeted space-time}
\hskip 0.6cm

    Intuitively, two-sheeted space-time can be viewed as an extension of the
physical four-dimensional space-time manifold $M$ by a discrete,
internal space $\{a,b \}$. In other words, it is a discretized
version of Kaluza-Klein theory \cite{KK} where the circle is
replaced by two discrete points in the fifth dimension. This
discretized Kaluza-Klein theory avoids the truncation
inconsistency problem of the original theory since it has a finite
field content with a finite mass spectrum \cite{VW2, VW1}.

As discussed in the preceding sections, the requisite input for
this model is the spectral triple, to which we apply the general
techniques discussed in Sects 2.3 and 2.4. The algebra of smooth
functions considered in Section 2.2 becomes $C^\infty(M,{\bf C})$
tensored with the complex-valued functions on the set $\{a,b \}$.
Clearly any such function, $f:\{a,b\} \rightarrow {\bf C}$ can be
written $f=f_a \oplus f_b$, each summand being isomorphic to ${\bf
C}$ itself.

\setcounter{equation}{1}
$$
{\cal A}\equiv C^\infty({\cal M}) \otimes ({\bf C} \oplus {\bf
C})\cong C^\infty({\cal M},{\bf C}) \oplus C^\infty({\cal M}, {\bf
C}). \eqno{(\thesection .\theequation)} \label{2S-ALGE}
$$

   In our previous work, where we were concerned with only gravity
   \cite{ LVW, VW1, VW2}, it was sufficient to have the algebra
$ C^\infty( R, {\cal M}) \bigoplus C^\infty( R, {\cal M})$. In the
present paper, since we are also interested in matter fields
coupled to gravity, we choose to begin with the more general
algebra (\ref{2S-ALGE}). However, we will find at the end, physics
determines a subalgebra of (\ref{2S-ALGE}) as physically relevant.

   In Eqn.(\ref{2S-ALGE}) we have essentially doubled the algebra presented
in Sect.2.2, it is natural to take the representation space as the
Hilbert space of square-integrable sections of a spinor bundle,

\refstepcounter{equation}
$$
{\cal H} \equiv L^2(S, {\cal M}) \oplus L^2(S, {\cal M}).
\eqno{(\thesection .\theequation)} \label{4.2}
$$

    The third element of the spectral triple is the generalized self-adjoint Dirac operator
\refstepcounter{equation}
$$
D = \Gamma^M D_M, ~~ M = 0,1,2,3, 5 \eqno{(\thesection
.\theequation)} \label{4.3}
$$

    with the generalization of the gamma matrices $\gamma^\mu$ given by
\refstepcounter{equation}
$$
\begin{array}{c}
      \Gamma^\mu
    \end{array} \equiv
\left(\begin{array}{cc}
      \gamma^\mu  &   0   \\
     0    &  \gamma^\mu
    \end{array}\right) ,\hspace{.5cm}
\begin{array}{c}
    \Gamma^5
    \end{array} \equiv
\left(\begin{array}{cc}
    0  & \gamma^5 \\
     \gamma^5  & 0
    \end{array}\right).
\eqno{(\thesection .\theequation)} \label{4.4}
$$

    With the basis specified, it is now possible to give an explicit form of the
generalized derivatives on the functions of the Hilbert space
${\cal H}$. The direct generalization of the ordinary spacetime
derivative $\partial_\mu$ follows obviously,

\refstepcounter{equation}
$$
\begin{array}{c}
    D_\mu \equiv
    \end{array}
\left(\begin{array}{cc}
    \partial_\mu & 0 \\
    0         & \partial_\mu
    \end{array}\right).
\eqno{(\thesection .\theequation)} \label{4.5}
$$

    These derivatives act along each of the space-time sheets. The derivative between the two sheets is given as
\refstepcounter{equation}
$$
D_5 \equiv \sigma \bar D_5, \eqno{(\thesection .\theequation)}
\label{4.6}
$$
where \refstepcounter{equation}
$$
\begin{array}{c}
      \bar D_5
    \end{array} \equiv
\left(\begin{array}{cc}
      0  &   m   \\
     m    &  0
    \end{array}\right) ,\hspace{.5cm}
\begin{array}{c}
    \sigma
    \end{array} \equiv
\left(\begin{array}{cc}
     0  & 1 \\
     1  & 0
    \end{array}\right).
\eqno{(\thesection .\theequation)} \label{4.7}
$$

    The parameter $m$ has the dimension of mass in order to give the
    fifth component of spacetime the same dimension as the other components.
    If we formally write the operator $DF$ as,

\refstepcounter{equation}
$$
D(F) \equiv [D, F] = \Gamma^M (DF)_M = \Gamma^M D_M F,
\eqno{(\thesection .\theequation)} \label{4.8}
$$
we indeed obtain the derivative along the fifth dimension with the
same dimension as other derivatives \refstepcounter{equation}
$$
D_5 F = \sigma[ \bar D_5, F] =  m(F -\tilde F), \eqno{(\thesection
.\theequation)} \label{4.9}
$$
where \refstepcounter{equation}
$$
\begin{array}{c}
    \tilde F
    \end{array} =
\left(\begin{array}{cc}
       f_2    &     0 \\
       0        &  f_1
    \end{array}\right).
\eqno{(\thesection .\theequation)} \label{4.10}
$$

    As required for the spectral triple, the operator $D$ on ${\cal H}$ is
self-adjoint with respect to the inner product
\refstepcounter{equation}
$$
\lacute \psi |D\phi \racute = \lacute D^\dagger\psi |\phi \racute
~~~~~~ , ~~~~~~ D=D^\dagger. \eqno{(\thesection .\theequation)}
\label{4.11}
$$

    A straightforward definition of the generalized wedge product used in the
previous work \cite{LVW, VW2, VW1} is the fully anti-symmetric
product that truncates all the ``junk'' forms in a trivial way,
namely,
 \refstepcounter{equation}
$$
\Gamma^M \wedge \Gamma^N = {1\over 2}(\Gamma^M \Gamma^N - \Gamma^N
\Gamma^M). \eqno{(\thesection .\theequation)} \label{WEDGEMN}
$$

        However, in this paper, since we would like to have the quartic Higgs potential
in the gauge sector, we will adopt the wedge product with the
following change in Eqn.(4.12), when $ M,N =5$,
\refstepcounter{equation}
$$
\Gamma^5 \wedge \Gamma^5 = 1. \eqno{(\thesection .\theequation)}
\label{WEDGE55}
$$

    This extended definition of the wedge product can be used in noncommutative geometric
    constructions of the Standard Model. The emergence of a quartic potential, however, as we shall see
    later, requires the fifth dimension to be complex.

    With the given spectral triple, it is straightforward to construct the algebra of forms on the two-sheeted
space-time. For later use, we work out the generalized
$\gamma$-matrix representations of one-forms and their
derivatives. The exterior derivative of any odd form is its
anti-commutator with the operator $D$ as a consequence of the
Leibnitz rule for the generalized forms (2.\ref{Leibnitz-pform}).
Hence the exterior derivative $DU$ of the generalized one-form
$U=\Gamma^M U_M$ is

\refstepcounter{equation}
$$
DU = \{D, U\} = \Gamma^M \wedge \Gamma^N (DU)_{MN},
\eqno{(\thesection .\theequation)} \label{4.14}
$$
    The components $(DU)_{MN}$ can be calculated with the help of the formulae
(4.\ref{WEDGEMN}) and (4.\ref{WEDGE55}) giving
\refstepcounter{equation}
$$
\begin{array}{ccc}
(DU)_{\mu \nu} & = & {1 \over 2} (D_\mu U_\nu - D_\nu U_\mu)\\
(DU)_{\mu 5} & = & - (DU)_{5 \mu} = {1 \over 2 }( D_\mu U_5 - D_5
U_\mu )\\
(DU)_{55} & = & m ( U_5 + \tilde U_5).
\end{array}
\eqno{(\thesection.\theequation)} \label{GENDEVCOMP}
$$

\subsection{ Riemannian geometry on a two-sheeted space-time}

The Equivalence Principle extended to our two-sheeted space-time
requires a locally orthonormal basis. Continuing to work with the
$\Gamma$-representation, the locally orthonormal $\Theta^A$ given
in Sect.3 are now represented by $\Gamma^A$,

\refstepcounter{equation}
$$
\begin{array}{c}
      \Gamma^a
    \end{array} \equiv
\left(\begin{array}{cc}
      \gamma^a  &   0   \\
     0    &  \gamma^a
    \end{array}\right) ,\hspace{.5cm}
\begin{array}{c}
    \Gamma^{\dot 5}
    \end{array} \equiv
\left(\begin{array}{cc}
    0  & \gamma^5 \\
     \gamma^5  & 0
    \end{array}\right),
\eqno{(\thesection .\theequation)} \label{2S-GREP}
$$
where $\gamma^a$ and $\gamma^5$ are the usual flat Dirac matrices.
By choosing $\gamma^5$ we have specialized to the two sheets of
spacetime of chiral spinors. We shall use a ${\dot 5}$ index in
the orthonormal basis to distinguish it from $5$ in the general
case.

   The inner product now can be also taken as a trace over the Clifford indices. In the
representation (4.\ref{2S-GREP}), the following orthogonality is
manifest:

\refstepcounter{equation}
$$
< \Gamma^A, \Gamma^B> ~~=~~ Tr (\Gamma^A \Gamma^B )~~=~~ 2
\eta^{AB}. {\bf 1} \eqno{(\thesection .\theequation)} \label{4.17}
$$

   The curvilinear $\Theta^M$ will be represented by $\Gamma^M$.
   To obtain $\Gamma^M$ and define the metric, we postulate the
   generalized vielbeins $E^M_A$, as the following diagonal matrix zero-forms:

$$
\begin{array}{c}
     E^\mu_a
    \end{array} \equiv
\left(\begin{array}{cc}
    e^\mu_{1 a} &  0 \\
     0    & e^\mu_{2 a}
    \end{array}\right) ,\hspace{.5cm}
\begin{array}{c}
    E^\mu_{\dot 5}
    \end{array} \equiv
\begin{array}{c}
     0
    \end{array},
$$

$$
\begin{array}{c}
    E^{5}_a
    \end{array} \equiv -
\left(\begin{array}{cc}
    a_{1a} &  0 \\
     0    & a_{2a}
    \end{array}\right) \equiv
\begin{array}{c}
    - A_a = - E^\mu_a A_\mu
    \end{array} ,
$$
\refstepcounter{equation}
$$
\begin{array}{c}
    E^{5}_{\dot 5}
    \end{array} \equiv
\left(\begin{array}{cc}
    \phi_1^{-1} &  0 \\
    0     & \phi_2^{-1}
    \end{array}\right) \equiv
\begin{array}{c}
    \Phi^{-1}
    \end{array},
\eqno{(\thesection .\theequation)} \label{VIELBEIN}
$$
where $e^\mu_{1,2a}$ are two different vielbeins on the two sheets
of space-time. Similarly, $ a_{1,2}$ and $\phi_{1,2}$ are
respectively vector and scalar fields. The dependence on the
space-time variable ,x, of the above vielbeins is understood. In
what follows, we shall, for the sake of simplicity of the
formulas, continue to suppress the x-dependence.

     The vielbeins $E^M_A$ are invertible giving,
$$
\begin{array}{c}
     E^a_\mu
    \end{array} \equiv
\left(\begin{array}{cc}
    e^a_{1 \mu} &  0 \\
     0    & e^a_{2 \mu}
    \end{array}\right) ,\hspace{.5cm}
\begin{array}{c}
    E^a_{5}
    \end{array} \equiv
\begin{array}{c}
     0
    \end{array},
$$
\refstepcounter{equation}
$$
\begin{array}{c}
    E^{{\dot5}}_\mu
    \end{array} \equiv
\begin{array}{c}
    A_\mu \Phi
    \end{array} , \hspace{.5cm}
\begin{array}{c}
    E_{5}^{\dot 5}
    \end{array} \equiv
\left(\begin{array}{cc}
    \phi_1 &  0 \\
    0     & \phi_2
    \end{array}\right) \equiv
\begin{array}{c}
    \Phi
    \end{array}.
\eqno{(\thesection .\theequation)} \label{VIELBEIN-INV}
$$
    Then, the $\Gamma^M$ matrices are given as
$
  \Gamma^M = \Gamma^A E^M_A,
$

\refstepcounter{equation}
$$
\begin{array}{c}
      \Gamma^\mu
    \end{array} \equiv
\left(\begin{array}{cc}
    \gamma^a  e^\mu_{1a}  &   0   \\
     0    &  \gamma^a e^\mu_{2a}
    \end{array}\right) ,\hspace{.5cm}
\begin{array}{c}
    \Gamma^5
    \end{array} \equiv
\left(\begin{array}{cc}
    - a_1  & \gamma^5 \phi_2^{-1} \\
     \gamma^5 \phi_1^{-1}  & - a_2
    \end{array}\right),
\eqno{(\thesection .\theequation)} \label{GAMMAM}
$$
where $ a_{1,2} = \gamma^a a_{1,2 a} $ are two gauge connection
one-forms on the two sheets.

     Using the inner product defined as the trace on the Hilbert space, we can calculate the
representation of the metric tensor as follows:
\refstepcounter{equation}
$$
\begin{array}{ccc}
G^{MN} ~~&=&~~ {1 \over 2}Tr(\Gamma^M \Gamma^N)=~~{ 1 \over 2}
{\tilde E}^M_A
Tr(\Gamma^A \Gamma^B) E^N_B\\
~~&=&~~ {\tilde E}^M_A \eta^{AB} E^N_B.
\end{array}
 \eqno{(\thesection
.\theequation)} \label{2S-METRIC}
$$

In terms of the functions $E^\mu_a$, $A_\mu$ and $\Phi$ the metric
tensor can be expressed as,
\refstepcounter{equation}
$$
\begin{array}{ccccc}
G^{\mu\nu} & = & G_0^{\mu \nu} &,& \hskip 5mm G^{\mu 5} =
- A^\mu,\\
G^{5\mu} & = & - A^\mu & , & \hskip 5mm G^{ 5 5} = A^2 +
\Phi^{-2},
\end{array} \eqno{(\thesection .\theequation)}
\label{4.22}
$$
\refstepcounter{equation}
$$
\begin{array}{ccccc}
G_{\mu\nu} & = & G_{0\mu \nu} + A_\mu A_\nu \Phi^2
& , & G_{\mu 5} = A_\mu \Phi^2,\\
G_{ 5 \mu} & = & A_\mu \Phi^2 & , & G_{ 5 5} = \Phi^2.
\end{array}
\eqno{(\thesection .\theequation)} \label{4.23}
$$

    The function $G_0$ is a generalized function with the metrics of the
two space-time sheets on the diagonal \refstepcounter{equation}
$$
\begin{array}{c}
    G_0^{\mu\nu} \equiv
    \end{array}
\left(\begin{array}{cc}

    e^\mu_{1a} \eta^{ab} e^\nu_{1b} & 0 \\
    0         & e^\mu_{2a} \eta^{ab} e^\nu_{2b}
    \end{array}\right).
\eqno{(\thesection .\theequation)} \label{4.24}
$$

As we see from the trace of the generalized gamma matrices, no
restrictions on the vielbeins can be derived if a consistent
choice of basis is selected. We can only derive the formulae of
generalized metric components in terms of the given vielbein
components from Eqn.(4.\ref{2S-METRIC}), which is in perfect
analogy with the usual Riemannian geometry.

 In terms of $\Gamma$-representations, an arbitrary
generalized one-form $U$ can be represented in both $\Gamma$-bases
as follows:

$$
U = \Gamma^A U_A = \Gamma^M U_M,
$$
where $U_M$ and $U_A$ are diagonal matrix functions
\refstepcounter{equation}
$$
\begin{array}{c}
      U_A
    \end{array} =
\left(\begin{array}{cc}
    u_{1A}  &   0   \\
     0    &  u_{2A}
    \end{array}\right) ,\hspace{.5cm}
\begin{array}{c}
    U_M
    \end{array} =
\left(\begin{array}{cc}
     u_{1M}  & 0 \\
     0  & u_{2M}
    \end{array}\right).
\eqno{(\thesection .\theequation)} \label{4.25}
$$

The functions $U_M$ and $U_A$ are related to each other by the
transformation laws,

\refstepcounter{equation}
$$
U_M = E^A_M U_A, \hskip 5mm U_A =E^M_A U_M. \eqno{(\thesection
.\theequation)} \label{1FORMTRANS0}
$$

In terms of the functions $E^a_\mu$, $A_\mu$ and $\Phi$, the
transformation laws can be written as
$$
U_\mu = E^a_\mu U_a + A_\mu \Phi U_{\dot 5}, \hskip 5mm U_5 = \Phi
U_{\dot 5},
$$
\refstepcounter{equation}
$$
U_a = E^\mu_a (U_\mu - A_\mu U_5 ) , \hskip 5mm U_{\dot 5} =
\Phi^{-1} U_5. \eqno{(\thesection .\theequation)}
\label{1FORMTRANS}
$$

The multiplication of two one-forms $U$ and $V$ is given by

\refstepcounter{equation}
$$
U V = (\Gamma^A U_A).(\Gamma^B V_B) = (\Gamma^M U_M).(\Gamma^N
U_N) \eqno{(\thesection .\theequation)}. \label{2S-1FORMPROD}
$$

   Here we want to address two issues concerning the multiplication of 1-forms:
   Firstly, as we don't know the explicit form of the matrices $\Gamma^M$ to rearrange
the order of terms in NCG, it is convenient to carry out the
multiplication in the $\Gamma^A$ basis. In fact, using the
explicit form of $\Gamma^A$ in Eqn.(4.\ref{2S-GREP}), we can
reorder the product in Eqn.(4.\ref{2S-1FORMPROD}). Secondly, as
the explicit forms of the basis $\Gamma^A$ are known, we can
define the wedge product by anti-symmetrizing the product except
for the component $\Gamma^{\dot 5} \wedge \Gamma^{\dot 5}$, which
we will keep as a formal non-vanishing basis of the 'two-forms'
algebra $\Omega^2({\cal A})$. That is to say, we have to define
the wedge product of the $\Gamma^A$ matrices with the basis
$\sigma^{AB}$ of the 'two-forms' module $\Omega^2({\cal A})$, to
be given by
\refstepcounter{equation}
$$
\sigma^{AB} = \Gamma^A \wedge \Gamma^B = {1 \over 2} [\Gamma^A,
\Gamma^B], \eqno{(\thesection .\theequation)}. \label{ANTI2FORM}
$$
when both $A$ and $B$ are not ${\dot 5}$. And when $A=B={\dot 5}$
we have
\refstepcounter{equation}
$$
\Gamma^{\dot 5} \wedge \Gamma^{\dot 5} \eqno{(\thesection
.\theequation)}. \label{4.30}
$$
as a formal basis vector. This component will generate a Higgs
potential for the gauge fields.

 From Eqn.(4.\ref{ANTI2FORM}), it is clear that
\refstepcounter{equation}
$$
\Gamma^A \wedge \Gamma^B = - \Gamma^B \wedge \Gamma^A.
\eqno{(\thesection .\theequation)}. \label{4.31}
$$
except the product $\Gamma^{\dot 5} \wedge \Gamma^{\dot 5}$.

  Now we know how to reorder the product in Eqn.(4.\ref{1FORMTRANS}) to define the
  wedge product of two
1-forms. The generalized functions commute with the matrices
$\Gamma^a$. The commutation rule of a function $F$ with the matrix
$\Gamma^{\dot 5}$ is as follows \refstepcounter{equation}
$$
F\Gamma^{\dot 5} = \Gamma^{\dot 5} \tilde F. \eqno{(\thesection
.\theequation)}. \label{4.32}
$$
It then follows that the product of two 1-forms $U$ and $V$ in
Eqn.(4.\ref{2S-1FORMPROD}) can be defined as the wedge product

\refstepcounter{equation}
$$
U \wedge V = \Gamma^A \wedge \Gamma^B ( U\wedge V)_{AB}.
\eqno{(\thesection .\theequation)}. \label{4.33}
$$

   and using the Eqns.(4.\ref{1FORMTRANS0}), we can compute the component of the wedge product
$U\wedge V$ in the locally flat frame,
\refstepcounter{equation}

$$
\begin{array}{ccl}
 (U \wedge V)_{ab} & = & {1 \over 2}( U_a V_b - U_b V_a)  ={1
\over 2}( E^\mu_a E^\nu_b - E^\nu_a E^\mu_b) \\
&~& ~~~\times(U_\mu V_\nu +  A_\mu(U_\nu V_5 - V_\nu U_5)),\\
(U \wedge V)_{{\dot 5}a} & = & -(U \wedge V)_{a {\dot 5}} =
{1\over 2}(U_{{\dot 5}}V_a - \tilde U_a V_{{\dot 5}})\\
&=&  {1 \over 2}\Phi^{-1}[ U_5 E^\mu_a (V_\mu - A_\mu V_5)
 - \tilde E^\mu_a V_5( \tilde U_\mu - \tilde A_\mu \tilde
U_5)], \\
(U \wedge V)_{{\dot 5}{\dot 5}}& = & \tilde U_{{\dot 5}} V_{{\dot
5}} ~~ = ~~ (\Phi \tilde \Phi)^{-1} \tilde U_5 V_5.
\end{array}
\eqno{(\thesection.\theequation)}. \label{WEDGE1}
$$

As a special case, we can calculate the wedge product of
$\Gamma^M$ one-forms as
\refstepcounter{equation}
$$
\begin{array}{ccc}
(\Gamma^M \wedge \Gamma^N)_{ab} & = &  {1 \over 2}(E^M_a E_b^N -
E_a^N E_b^M ),\\
(\Gamma^M \wedge \Gamma^N)_{{\dot 5} a} & = & {1 \over 2} (
E^M_{{\dot 5}} E^N_a - \tilde E^M_a  E^N_{{\dot 5}}),\\
(\Gamma^M \wedge \Gamma^N)_{{\dot 5}{\dot 5}} & = &\tilde
E^M_{{\dot 5}} E^N_{{\dot 5}}.
\end{array}
\eqno{(\thesection .\theequation)}. \label{4.37}
$$

The derivative $DU$ of the 1-form $U$ in the curved space-time
basis follows from (4.\ref{GENDEVCOMP}) and is given by

\refstepcounter{equation}
$$
\begin{array}{ccc}
DU & =&  \Gamma^M \wedge \Gamma^N (DU)_{MN} =  \Gamma^A \wedge
\Gamma^B (DU)_{AB} \\
&=&  \Gamma^A \wedge \Gamma^B (\Gamma^M \wedge \Gamma^N)_{AB}
(DU)_{MN} ,
\end{array}
\eqno{(\thesection .\theequation)}.
\label{4.38}
$$
and hence
\refstepcounter{equation}
$$
\begin{array}{ccc}
 (DU)_{ab} & = & {1\over 2}(E^\mu_a E^\nu_b - E^\nu_a
E^\mu_b)( D_\mu U_\nu + A_\mu (
 D_\nu U_5 - D_5 U_\nu),\\
 (DU)_{{\dot 5}a}& = & - (DU)_{a{\dot
5}} = - {1 \over 2} \Phi^{-1}[ {1 \over 2} (E^\mu_a + \tilde
E^\mu_a)(D_\mu U_5 - D_5U_\nu) \\
&~& + m( E^\mu_a A_\mu - \tilde E^\mu_a \tilde A_\mu) ( U_5 +
\tilde U_5) ],
\\
(DU)_{{\dot 5}{\dot 5}} &=& - m(\Phi \tilde \Phi)^{-1} (U_5 +
{\tilde U}_5)
\end{array}
\eqno{(\thesection .\theequation)}.\label{DEVCOMP}
$$
The inner product of two 2-forms \refstepcounter{equation}
$$
<\Gamma^A \wedge \Gamma^B, \Gamma^C \wedge \Gamma^D> \equiv {1
\over 4} Tr_4( \Gamma^A \Gamma^B \Gamma^C \Gamma^D) = \eta^{AD}
\eta^{BC} - \eta^{AC} \eta^{BD}. \eqno{(\thesection
.\theequation)}. \label{4.41}
$$

    Similarly, we can give a representation of the general p-forms, multiplication rules of
a p-form and a q-form and the derivative of general p-forms. The
use of the two frames is essential in such calculus. To summarize,
it is convenient to carry out the multiplication in the $\Gamma^A$
basis and the derivatives in the $\Gamma^M$ basis and then
transform to the $\Gamma^A$ to obtain the desired components. The
physical fields are given in the curved frame.

     The definition of $\Gamma^{{\dot 5}}\wedge \Gamma^{{\dot 5}}$ is rather arbitrary but
is required to retain the ${\dot 5}{\dot 5}$ component for a Higgs
potential to survive in the gauge theory. It is more natural to
introduce the discrete degrees of freedom as a pair of conjugate
derivatives $D_z , D_{{\bar z}}$ as in \cite{ZZbar,WIGNER} than
just a single $D_5$ in the Dirac operator.

   In the gravitational sector, the curvature arises from a set of connection
one-forms $\{\Omega^A_{B} ~ A,B = 0,...\dot 5 \}$. The connection
coefficients are defined as those functions which arise in the
expansion of $\Omega^A_B$ in the $\Gamma$-basis, as the
generalization of (3.\ref{COVDEV}), \refstepcounter{equation}
$$
\Omega^A_B \equiv \Gamma^C \Omega^A_{BC}. \eqno{(\thesection
.\theequation)} \label{OMEGA}
$$

\section{ Physical content of the gravity sector in two-sheeted space-time}
\hskip 0.6cm
     In Section 4., we have seen that the metric (vielbein) structure
of the two-sheeted space-time contains a tensor, a vector and a
scalar, which have generalized functions in
Eqn.(4.\ref{VIELBEIN}). This field content is exactly parallel to
the one in the traditional Kaluza-Klein theory \cite{KK}. However,
each generalized function has a pair of fields. Therefore, in
principle, the gravity sector in the two-sheeted space-time,
contains a pair of tensor, a pair of vector and a pair of scalar
fields.

     Intuitively as stated earlier, the two-sheeted space-time is a discretized
version of the general untruncated Kaluza-Klein theory \cite
{DUFF}, where the internal space circle is replaced by two points.
Therefore, intuition tells us that in general, we should expect
one member of each pair to be massless and the other massive \cite
{VW2}.

\subsection{ Minimal set of constraints and solutions of the
first Cartan structure equation}

     The essential point is how to impose a consistent system of
constraints to solve the generalized Cartan structure equations
without overconstraining the metric. The second Cartan structure
equation is used to calculate the curvature tensor in terms of the
connections. Therefore, it does not have much to do with
constraints. The first Cartan structure equation expresses a
relationship between connections and torsion. Its component
equations are as follows: \setcounter{equation}{0}
\refstepcounter{equation}
$$
\begin{array}{ccc}
  T_{abc}  & = & (D\Gamma_a)_{bc} + {1 \over 2}( \Omega_{abc} - \Omega_{acb})  \\
  T_{a{\dot 5}b} & = & (D\Gamma_a)_{{\dot5}b} + {1 \over 2}(\Omega_{a{\dot 5}b} - \Omega_{ab{\dot 5}}) \\
  T_{{\dot 5}ab} & = & (D\Gamma_{\dot 5})_{ab} + {1 \over 2}(\Omega_{{\dot 5}ab} - \Omega_{{\dot 5}ba}) \\
  T_{{\dot 5}{\dot 5}b} & = & (D\Gamma_{\dot 5})_{{\dot 5}b} +
           {1 \over 2}(\Omega_{{\dot 5}{\dot 5}b} - \Omega_{{\dot 5}b{\dot 5}}) \\
  T_{ab{\dot 5}} & = & (D\Gamma_a)_{b{\dot 5}}+
            {1 \over 2}(\Omega_{ab{\dot 5}} - \Omega_{a{\dot 5}b}) \\
  T_{{\dot 5} a {\dot 5}} & = & (D\Gamma_{{\dot 5}})_{a{\dot 5}}+
            {1 \over 2}(\Omega_{{\dot 5}a{\dot 5}} - \Omega_{{\dot 5}{\dot 5}a}) \\
  T_{a{\dot 5} {\dot 5}} & = & (D\Gamma_a)_{{\dot 5}{\dot 5}}+
            {1 \over 2}(\Omega_{a{\dot 5}{\dot 5}} + \tilde \Omega_{a{\dot 5}{\dot 5}}) \\
  T_{{\dot 5}{\dot 5}{\dot 5}} & = & (D\Gamma_{\dot 5})_{{\dot
  5}{\dot 5}} + {1 \over 2} ( \Omega_{{\dot 5}{\dot 5}{\dot 5}} + \tilde \Omega_{{\dot 5}{\dot 5}{\dot
  5}}).
\end{array}
\eqno{(\thesection .\theequation)} \label{TORCON}
$$
Obviously these equations are not enough to determine both the
torsion and the connection, even in the case of the ordinary
geometry.

     The torsion free condition $T^A=0$ together with the metric
compatibility condition turns out to be too restrictive giving
only a theory without massive excitations. However, it is possible
to impose constraints in such a way as to retain all the
postulated fields in the metric structure in more than one way
\cite{VW2, VW1, Vtorsion}. In this paper we will follow the
results of \cite{Vtorsion} to impose the following minimal set of
constraints:

i) Spacetime torsion free condition

\refstepcounter{equation}
$$
  T_{aBC} ~~~=~~~ 0 , ~~~
  \eqno{(\thesection.\theequation.a)}
$$
ii) Metric compatibility condition
$$
   \Omega_{AB} ~~~=~~~- \Omega_{BA}
  \eqno{(\thesection.\theequation.b)}
$$
iii) An additional condition on the connections.
$$
  \Omega_{AB {\dot 5}} ~~~=~~~  0
\eqno{(\thesection.\theequation.c)}
$$
    With these conditions, Eqn.(5.\ref{TORCON}) can be solved
giving the following solutions for the non-vanishing connection
and torsion-tensor components,
\refstepcounter{equation}
$$
\begin{array}{ccc}
  \Omega_{abc}  & = & -(D\Gamma_a)_{bc} + (D\Gamma_b)_{ac} - (D\Gamma_c)_{ba} \\
  \Omega_{a{\dot 5}b} & = & - 2 (D\Gamma_a)_{{\dot 5} b} \\
  \Omega_{{\dot 5}ab} & = & 2 (D\Gamma_a)_{{\dot 5} b} \\
  T_{{\dot 5}bc} & = & (D\Gamma_{\dot 5})_{bc} + (D\Gamma_b)_{{\dot 5} c} - (D\Gamma_c)_{{\dot 5} b} \\
  T_{{\dot 5}{\dot 5}b} & = & (D\Gamma_{\dot 5})_{{\dot 5}b} \\
  T_{{\dot 5}{\dot 5}{\dot 5}} & = & (D\Gamma_{\dot 5})_{{\dot
  5}{\dot 5}}.
\end{array}
\eqno{(\thesection .\theequation)} \label{SOLU}
$$

The righthand side of Eqns containing the components of $
(D\Gamma_A)$ can be expressed in terms of metric components (
vielbeins ) and their derivatives by the rules already stated.
Thus, we have non-vanishing torsion and connection coefficients in
terms of metric fields.

    Here let us note that no constraints are imposed on the
metric components (vielbeins). Therefore, all the six fields will
contribute to the scalar curvature. If all of them have
appropriate kinetic terms in the final action, they will represent
themselves as physical fields of the gravity sector of our theory.

   However, the fields $e^\mu_{1,2~a}$, $a_{1,2~\mu}$ and
$\phi_{1,2}$ turn out not to be good variables of the theory. They
do not give rise to mass eigenstates except in the case $e_1 =
e_2, a_1=a_2,$ when the theory reduces exactly to the zero mode
sector of the Kaluza-Klein theory with massless fields.

In the next subsection, we define linear combinations of these
fields which lead to appropriate fields leading to mass
eigenstates.

\subsection { New physical field variables}

    We begin with the following linear combinations as new physical field variables:
\refstepcounter{equation}
$$
\begin{array}{ccc}
    e^\mu_{\pm a} & = &{1 \over 2}( e^\mu_{1~a} \pm e^\mu_{2~ \nu}), \\
    a_{\pm \mu \nu} & = &{1 \over 2}( a_{1 \mu} \pm a_{2 \mu}), \\
    \phi_{\pm} & = & { 1 \over 2} (\phi_1 \pm \phi_2).
\end{array}
\eqno{(\thesection .\theequation)} \label{5.4}
$$

  When $e^\mu_{-~\nu} = a_{-~\mu} = \phi_{-} =0
  $, the theory reduces to the usual Kaluza-Klein theory with $e^\mu_{+
~a}$ as the vierbein, $a_{+ \mu}$ the vector field and $\phi_+$
the Brans-Dicke scalar.

To obtain the "physical" action functional in terms of the new
variables, we need to define $E^\mu_a$ and its inverse $E^a_\mu$
in terms of the new variables. This is not as straightforward as
one would think.

To this end, we start with \refstepcounter{equation}
$$
E^\mu_a = \left( \begin{array}{cc}
                 e^\mu_{1a} & 0 \\
                 0   &  e^\mu_{2a}
                 \end{array}
              \right)
           =  e^\mu_a{\bf 1} + v^\mu_a.{\bf r}~~~~,
\eqno{(\thesection .\theequation)} \label{5.5}.
$$
where  $ v^\mu_a = e^\mu_{-a} $ and $ e^\mu_a = e^\mu_{a+}$. Let
$e^a_\mu = (e^\mu_a)^{-1}$ and
\refstepcounter{equation}
$$
\begin{array}{c}
      {\bf 1}
    \end{array} =
\left(\begin{array}{cc}
      1  &   0   \\
      0   &  1
    \end{array}\right) ,\hspace{.5cm}
\begin{array}{c}
    {\bf r}
    \end{array} =
\left(\begin{array}{cc}
     1  &  0 \\
     0  & -1
    \end{array}\right).
\eqno{(\thesection .\theequation)} \label{5.6}.
$$
From the orthogonality relation of the vielbeins
\refstepcounter{equation}
$$
E^\mu_a E^b_\mu = \delta^a_b { \bf 1}
\eqno{(\thesection.\theequation)} \label{5.7}
$$
we obtain
\refstepcounter{equation}
$$
E^a_\mu = ( e^a_\mu - v^\nu_\mu s^a_\nu) { \bf 1} + s^a_\mu {\bf
r},
 \eqno{(\thesection.\theequation)} \label{5.8}
$$
where $s^a_\mu$ is a non-linear function of $v^a_\mu$ satisfying
\refstepcounter{equation}
$$
v^\nu_\mu + s_\sigma^\nu ( \delta_\mu^\sigma - v^\lambda_\mu
v_\lambda^\sigma ) = 0,
 \eqno{(\thesection.\theequation)}
\label{5.9}
$$
and $s^\nu_\sigma = e^\nu_a s^a_\sigma $.

  The tensor field $v^{\mu \nu}$ is a candidate for a massive excitation of
  the massless symmetric metric tensor.

 We can also use the metric tensors $g^{\mu \nu}$ and $g_{\mu \nu}$ to raise and lower indices and
 express the tensor fields $v$ and $s$ in a matrix notation, $v = [v^\nu_\mu]~~~,~~~
s = [s^\nu_\mu]$.

The following algebraic relations are also valid
 \refstepcounter{equation}
$$
vs= sv  ~~~,~~~  s = v ( v^2-1)^{-1}.
\eqno{(\thesection.\theequation)} \label{5.10}
$$

The usual space-time component of the generalized vierbein
$E^\mu_a$ and its inverse $E^a_\mu$ are determined completely in
terms of the usual vierbeins $e^\mu_a$, $e^a_\mu$ and the tensor
field $v^{\mu \nu}$.

Furthermore, we can define the vector and scalar components of the
metric by

\refstepcounter{equation}
$$
\begin{array}{ccc}
    A_{\mu}& = & a_{+~ \mu}{\bf 1} + a_{-~ \mu}{\bf r} ~~~~, \\
    \Phi & = & \phi_+ {\bf 1} + \phi_- {\bf r}~~~ .
\end{array}
\eqno{(\thesection .\theequation)} \label{5.11}
$$

   Hence, finally, the gravity sector of the theory contains $
e^\mu_a, e_\mu^a,$ the tensor field $ v^\mu_\nu$, the vector and
scalar fields $ a_{\pm \mu}, \phi_{\pm} $. The tensor field $
s^\mu_\nu $ is not an independent field. It is a non-linear
combination of $ v^\nu_\mu $. As we will see in the final
Lagrangian, the tensor field $v^{\mu \nu}$ will be a mass
eigenstate. We are, therefore, led to make the assumption, $v^{\mu
\nu}$ = $v^{\nu\mu}$, in order to have $ v^{\mu \nu}$ as an
independent tensor field.

  As we shall also see, the signature of the kinetic terms of the
fields $v^{\mu \nu}$ , $ a_{\mu}$ and $\phi_-$ in the final
Lagrangian built from curvature and torsion terms of our theory,
requires the following field redefinitions,
\refstepcounter{equation}
$$
\begin{array}{ccc}
   v^a_\mu  &~\longrightarrow~& i v^a_\mu , \\
   a_{-\mu} &~\longrightarrow~& i a_{-\mu},  \\
   \phi_-     &~\longrightarrow~& i \phi_-.
\end{array}
\eqno{(\thesection .\theequation)} \label{FIREDEF1}
$$
  This field redefinition will be used hereafter in this paper.

  It is convenient to introduce the following quantities in further
calculations: \refstepcounter{equation}
$$
\begin{array}{ccc}
P_{a\mu\nu} &=&  \partial_\mu(e_{a\nu} +
s_{a\lambda} v^\lambda_\nu) - 2m a_{\nu -} s_{a\mu}\\
Q_{a\mu\nu} &=& \partial_\mu s_{a\nu} + 2m a_{\nu+} s_{a\mu}\\
X^{[\mu, \nu]}_{bc} &~=~& ( e^\mu_b
e^\nu_c - e^\nu_b e^\mu_c + v^\nu_b v^\mu_c - v^\mu_b v^\nu_c ) \\
Y^{[\mu, \nu]}_{bc} &~=~&  ( e^\mu_b v^\nu_c - e^\nu_b v^\mu_c + v^\mu_b e_c^\nu - v^\nu_b e^\mu_c ). \\
\end{array}
\eqno{(\thesection .\theequation)} \label{5.13}
$$
  In several formulae of this theory, $P_{a\mu\nu}$ and $Q_{a\mu \nu}$
generalize the field derivatives $\partial_\mu e_{a\nu}$ and
$\partial_\mu v_{a\nu}$. The $X^{[\mu, \nu]}_{ab}$ generalize the
usual term $e^\mu_a e^\nu_b - e^\nu_a e^\mu_b$ which is common in
general relativity. Some useful contraction formulae for $X^{[\mu,
\nu]}_{ab}$ and $Y^{[\mu, \nu]}_{ab}$ can be found in Appendix
A.2. These formulae allow us to write the action in a compact
form.

   For calculational convenience, it is useful to introduce the
following projection operations for the component $f_\pm$ of a
generalized function $F$

\refstepcounter{equation}
$$
\begin{array}{ccccc}
    (F)_+ & = &  f_+ & = & {1 \over 2} Tr(F) ~~~,~ \\
    (F)_- & = & f_- & = & {1 \over 2} Tr(F. {\bf r})~.~~
\end{array}
\eqno{(\thesection .\theequation)} \label{PROJ1}
$$

The following relations are also useful in practical calculations
of products of generalized functions

\refstepcounter{equation}
$$
\begin{array}{ccc}
(F.G)_+ =  (F)_+ (G)_+ + (F)_- (G)_-, \\
(F.G)_- =  (F)_+ (G)_- + (F)_- (G)_+ .
\end{array}
\eqno{(\thesection .\theequation)} \label{PROJ2}
$$

\subsection{Curvature, Torsion and Lagrangian}

The second structure equation is generalized in the following
form,
 \refstepcounter{equation}
$$
 R_{AB} = D\Omega_{AB} + \eta^{CD} \Omega_{AC} \wedge
 \Omega_{DB}.
\eqno{(\thesection .\theequation)} \label{CURV}
$$

   With the connections known from Eqn.(5.\ref{SOLU}), we can calculate all
the components of the curvature in Eqn.(5.\ref{CURV}). However, we
are interested in only few components that contribute to the
generalized scalar curvature

\refstepcounter{equation}
$$
\begin{array}{ccc}
  R & = & {1\over 2} Tr <\Gamma^A \wedge \Gamma^B, R_{AB} > =  R_{abcd +} \eta^{ad} \eta^{bc} + 2 R_{a{\dot 5}{\dot
  5}b+} \eta^{ab} \\
   & = & (D\Omega_{ab})_{cd+} \eta^{ad} \eta^{bc}
  + ( \Omega_{ae} \wedge \Omega_{bf})_{cd +} \eta^{ad} \eta^{cf} \eta^{be}\\
  &~& + 2 (D\Omega_{a{\dot 5}})_{{\dot 5}b+}\eta^{ab}
  + 2 (\Omega_{a{\dot 5}} \wedge \Omega_{{\dot 5}b})_{cd+} \eta^{ad}
  \eta^{bc}.
\end{array}
\eqno{(\thesection .\theequation)} \label{SCURV}
$$

The Lagrangian from curvature is defined as

\refstepcounter{equation}
$$
   {\cal L}_R =  {1 \over 16 \pi G_N} R = {\cal L}_{R1} + {\cal
   L}_{R2} + {\cal L}_{R3},
\eqno{(\thesection .\theequation)} \label{5.20}
$$
where
\refstepcounter{equation}
$$
  \begin{array}{ccc}
{\cal L}_{R1} & = & ( 16 \pi G_N)^{-1} (D\Omega_{ab})_{cd+} \eta^{ad} \eta^{bc} \\
{\cal L}_{R2} & = &  (16 \pi G_N)^{-1} (\Omega_{ae} \wedge \Omega_{bf})_{cd +} \eta^{ad} \eta^{cf} \eta^{be}\\
{\cal L}_{R3} & = &  (8 \pi G_N)^{-1} ((D\Omega_{a{\dot
5}})_{{\dot 5}b+}\eta^{ab}
   + (\Omega_{a{\dot 5}} \wedge \Omega_{{\dot 5}b})_{cd+} \eta^{ad}
  \eta^{bc}),\end{array}
\eqno{(\thesection .\theequation)} \label{5.21}
$$
where $G_N$ is the Newton gravitational constant. The detailed
calculations of the above terms are given in Appendix A.3.

  Examination of these results shows that the``curvature''
  Lagrangian contains the usual four-dimensional curvature (
  Einstein-Hilbert term) along with a full Lagrangian for the
  massive tensor field $v_{\mu \nu}$. It also contains mass and
  interaction terms for the scalar and vector components of the
  metric. However, it does not contain kinetic terms for the
  latter. In order to obtain the complete Lagrangian, we need to
  take into account the contribution from the non-vanishing
  components of the torsion. This is given by

\refstepcounter{equation}
$$
\begin{array}{ccc}
{\cal L}_T &~=~& -( 16 G^2_V)^{-1} Tr <T_A, T^A> ~=~  ( 8 G^2_V
)^{-1} Tr({\tilde
T}_{ABC} T^{ABC})~~~~~~~~~~~\\
&~=~&  ( 4 G^2_V )^{-1} ( T_{ABC+} T^{ABC}_+ - T_{ABC-}T^{ABC}_-)
~~~~~~~~~~~~~~~~~~~~\\

&~=~&  ( 4 G^2_V )^{-1} (T_{{\dot 5}bc+} T^{{\dot 5} bc}_+
  - T_{{\dot 5}bc-} T^{{\dot 5}bc}_-
  + 2T_{{\dot 5}{\dot 5}b+} T^{{\dot 5}{\dot 5} b}_+
  - 2T_{{\dot 5}{\dot 5}b-} T^{{\dot 5}{\dot 5} b}_-
   +~~ T_{{\dot 5}{\dot 5}{\dot 5} +}
  T^{{\dot 5}{\dot5}{\dot 5}}_+) \\
&~=~&  {\cal L}_{T1} + {\cal L}_{T2} +{\cal L}_{T3},
\end{array}
\eqno{(\thesection .\theequation)} \label{5.23}
$$

where the Lagrangians ${\cal L}_{T1}, {\cal L}_{T2}$ and $ {\cal
L}_{T3}$ are given in Appendix A.4. $G_V$ is a constant with
dimension of mass.

   Finally, the Einstein-Hilbert-Cartan Lagrangian for the gravity sector
in our theory is

\refstepcounter{equation}
$$
\begin{array}{ccc}
   {\cal L}_{E-H-C} &~=~&  {\cal L}_R + {\cal L}_T ~~~~~=~~ (16 \pi
   G)^{-1} Tr R + ( 16 G^2_V )^{-1} Tr < T_A, T^A> \\
   &~=~& {\cal L}_{R1} + {\cal L}_{R2} + {\cal L}_{R3} + {\cal L}_{T1} + {\cal L}_{T2} + {\cal L}_{T3}
\end{array}
   \eqno{(\thesection .\theequation)} \label{LAGRAV}
$$

The kinetic terms of the fields $v^a_\mu$, $a_{\mu-}$ and $
\phi_-$ in the resulting Lagrangian require the field
redefinitions defined in Eqn.(5.\ref{FIREDEF1}).

\section{ Matter fields in curved two-sheeted space-time}

With the curved space-time described in the previous section as
the background, we now proceed to construct Lagrangians for the
gauge and fermionic sectors. The gauge sector will consist of two
Abelian gauge fields with two Higgs scalar fields as part of a
generalized one-form. The fermionic sector will consist of a
left-chiral field on one sheet and a right-chiral field on the
other.

\subsection{ The gauge sector}

   Let the one-form, $U$, be specialized to the
   gauge and Higgs fields to be associated with the chiral
   fermions,
\setcounter{equation}{0}

$$
U = B = \Gamma^M(x) B_M,
$$
where

\refstepcounter{equation}
$$
\begin{array}{c}

B_\mu = \left(
\begin{array}{cc}
       b_{1\mu} & 0 \\
       0   & b_{2\mu}
\end{array}
\right) ~~,~~

B_5 = \left(
\begin{array}{cc}
       h_1 & 0 \\
       0   & h_2
\end{array}
\right),
\end{array}
\eqno{(\thesection .\theequation)} \label{6.1}
$$

or

$$
B_\mu = b_{\mu+} {\bf 1} + b_{\mu-} {\bf r} ~~,~~ B_5 = h_+{\bf 1}
+ h_- {\bf r},
$$

where

\refstepcounter{equation}
$$
    b_{\pm \mu} = {1 \over 2}( b_{1\mu} \pm b_{2\mu}) ~~,~~
    h_{\pm} = { 1 \over 2} ( h_1 \pm h_2).
\eqno{(\thesection .\theequation)} \label{6.2}
$$

Then, the curvature or field strengths are given by the two-form (
henceforth the $x$-dependence will be understood),

$$
    G  = DB + B \wedge B,
$$
with components, following
Eqns.(4.\ref{WEDGE1})-(4.\ref{DEVCOMP}),

\refstepcounter{equation}
$$
G_{ab} = {1 \over 2}( E^\mu_a E^\nu_b- E^\nu_a E^\mu_b) [ D_\mu
B_\nu  + A_\mu (D_\nu B_5 - D_5 B_\mu)]~, \eqno{(\thesection
.\theequation)} \label{STRE1}
$$
\refstepcounter{equation}
$$
\begin{array}{ccc}
 G_{a \dot 5}&=&{1 \over 2} E^5_{\dot 5}
 [ {1 \over 2} ( {\tilde E}^\mu_a + E^\mu_a ) (D_\mu B_5 -D_5 B_\mu) +  m (E^\mu_a
A_\mu - {\tilde E}^\mu_a {\tilde A}_\mu)( B_5 + {\tilde B}_5 ) \\
 &~& + ( {\tilde E}^\mu_a {\tilde B}_\mu - E^\mu_a B_\mu +  E^\mu_a
 A_\mu B_5 - {\tilde E}^\mu_a {\tilde A}_\mu {\tilde B}_5)B_5 ] ~~=~~ - G_{a{\dot 5}},
\end{array}
\eqno{(\thesection .\theequation)} \label{6.4}
$$

\refstepcounter{equation}
$$
G_{{\dot 5}{\dot 5}} = {\tilde E}^5_{\dot 5} E^5_{\dot 5} (
m({\tilde B}_5 + B_5) + {\tilde B}_5 B_5).
\eqno{(\thesection.\theequation)} \label{STRE3}
$$

In terms of the above components, the Lagrangian is given by
\refstepcounter{equation}
$$
{\cal L}_G = {1 \over 2g^2} < G , G > = - {1 \over g^2}({\tilde
G}_{ab} G^{ab} + 2 {\tilde G}_{a \dot 5} G^{a \dot 5} + { 1 \over
2} {\tilde G}_{\dot 5 \dot 5} G^{\dot 5 \dot 5})  ~.
\eqno{(\thesection.\theequation)} \label{LAGAU}
$$

  As in the case of the torsion Lagrangian in the gravity sector,
the gauge Lagrangian contains the kinetic terms for the fields
$b_{-\mu}$ and $h_-$ with a sign that requires a similar field
redefinition. Therefore, we introduce the following new field
variables in order to bring the gauge Lagrangian into a form,
which is consistent with the standard model with two abelian gauge
fields and a Higgs scalar.
\refstepcounter{equation}
$$
\begin{array}{ccc}
b_{\mu-} &~\leftrightarrow~& -i  b_{\mu-}, \\
\eta &~ \leftrightarrow ~& h_+ + m + h_- ,\\
\bar \eta &~\leftrightarrow~& h_+ + m - h_-,\\
b_{\mu \nu+} &~=~& \partial_\mu b_{\nu+}- \partial_\nu b_{\mu+}, \\
b_{\mu \nu-} &~=~& \partial_\mu b_{\nu-}- \partial_\nu b_{\mu-},\\
{\cal D}_\mu \eta &~=~&\partial_\mu \eta - 2ib_{\mu-}\eta,\\
 {\cal D}_\mu \bar \eta &~=~& \partial_\mu \bar \eta + 2 i b_{\mu-} \bar
\eta.
\end{array}
\eqno{(\thesection.\theequation)} \label{FIREDEF2}
$$
Substituting the even and odd projection of $G_{AB}$ from
Eqns.(6.\ref{STRE1})-(6.\ref{STRE3}) in Eqn.(6.\ref{LAGAU}), we
obtain the gauge sector Lagrangian ${\cal L}_G$. We write it in
the form
\refstepcounter{equation}
$$
{\cal L}_G = {\cal L}_{G1} + {\cal L}_{G2} + {\cal L}_{G3} + {\cal
L}_{G4}, \eqno{(\thesection.\theequation)}\label{LAGAU2}
$$
where ${\cal L}_{G1}$ contains the kinetic terms of the vector
gauge fields $b_{\mu+}$ and $ b_{\mu-}$ and ${\cal L}_{G2}$
contains the kinetic terms of the Higgs field. ${\cal L}_{G3} $
contains the quartic Higgs fields. ${\cal L}_{G4}$ contains the
remaining interaction terms of these fields with the scalar and
vector fields of the gravity sector.

   The Lagrangian (6.\ref{LAGAU2}), therefore, represents a generalization of the
standard model gauge sector involving two abelian gauge fields
coupled with a complex scalar with a quartic potential in the
curved two-sheeted space-time. More discussion about the physical
contents of this Lagrangian will be give in Sect.7.

\subsection{ Fermion sector}

We begin with the conventional Dirac Lagrangian in curved
space-time generalized to our two-sheeted space-time given by

\refstepcounter{equation}
$$
\begin{array}{ccc}
{\cal L}_F &~=~& i  \bar \Psi \Gamma^A ( E^M_A(D_M + i B_M) + {i
\over 4} \Gamma^B \Gamma^C \Omega_{BCA}) \Psi \\

 &~=~&  i  \bar \Psi \Gamma^a [ E^\mu_a(D_\mu + iB_\mu)+ {i
 \over 4} \Gamma^b \Gamma^c \Omega^{(0)}_{bca} + {i \over 2}
\Gamma^{\dot 5}
\Gamma^c \Omega_{{\dot 5}ca})\\
&~~~& +{i\over 4} \Gamma^a\Gamma^b\Gamma^c \Omega^{(1)}_{bca} + (
\Gamma^{{\dot 5}}\Phi^{-1} - \Gamma^a
E^\mu_aA_\mu)(D_5+ i B_5)]\Psi \\
&~=~& {\cal L}_{F1}+ {\cal L}_{F2} + {\cal L}_{F3} + {\cal
L}_{F4},
\end{array}
\eqno{(\thesection.\theequation)} \label{LAFE}
$$

where
$$
\Psi =  \left( \begin{array}{c}
              \psi_L \\
              \psi_R
              \end{array}
        \right),
$$

$$
\psi_L = {(1 + \gamma_5) \over 2} \psi ~~,~~ \psi_R = {(1-
\gamma_5)\over 2} \psi,
$$
 $\psi$ is the 4-component Dirac spinor and $ \Omega_{BCA}$ are the
 connection coefficients. The expressions for the non-vanishing
 $\Omega_{ABC}$ ( $\Omega_{abc}$, $\Omega_{a{\dot 5} b}$ ) are given
 in Eqn.(5.\ref{SOLU}). In terms of metric component fields and their
 derivatives, they are

$$
\Omega_{abc} = \Omega^{(0)}_{abc} + \Omega^{(1)}_{abc},
$$
where

\refstepcounter{equation}
$$
\begin{array}{ccc}
\Omega^{(0)}_{abc} & = & {1 \over 2}( \eta_{ad} E^\mu_b E^\nu_c -
\eta_{bd} E^\mu_a E_c^\nu + \eta_{cd} E_a^\mu E^\nu_b)( D_\mu E^d_\nu - D_\nu E^d_\mu),\\
\Omega^{(1)}_{abc} & = & {1 \over 2}(\eta_{ad} E^\mu_b E_c^\nu -
\eta_{bd} E^\mu_a E^\nu_c + \eta_{cd} E^\mu_a E_b^\nu) ( A_\nu D_5
E^d_\mu - A_\mu D_5 E^d_\nu ),
\end{array}
\eqno{(\thesection.\theequation)}\label{6.14}
$$

\refstepcounter{equation}
$$
\Omega_{a {\dot 5}b} = - \Omega_{{\dot 5} ab} = 2 \eta_{ad}
E^5_{\dot 5} E^\mu_b ( D_5 E^d_\mu).
\eqno{(\thesection.\theequation)}\label{6.15}
$$

The separation of $\Omega_{abc}$ into $\Omega^{(0)}_{abc}$ and $
\Omega^{(1)}_{abc} $ is  motivated by the consideration of the
fact that $\Omega^{(0)}$ contains only the metric vierbeins and
their generalizations to the two-sheeted space-time, where as
$\Omega^{(1)}$ contains the additional vector and scalar fields of
the gravity sector.

 The results of calculations of the Lagrangian ${\cal L}_F$ in
Eqn.(6.\ref{LAFE}) are given in Appendix A.6.

\section{Physical implications}
\setcounter{equation}{0}

   So far, we have concentrated on developing the mathematical
formalism within the framework of Conne's NCG. The various fields
we have introduced do not have the desired physical dimensions.
Just as we needed to redefine the fields ( Eqns.(5.\ref{FIREDEF1})
and (6.\ref{FIREDEF2}) ) to secure the coorect signs for the
kinetic terms, we need to rescale them using the avialable
dimensional parameters in our theory. To this end, we note that we
have only three dimensional parameters, $G_N$ the Newton
gravitational constant, $G_V$ the new gravitational constant from
the torsion and the parameter $m$ with dimensions of mass. The
dimensionless gauge coupling $g$ is the only other free parameter.
By fixing the standard coefficients of the kinetic terms, we
determine the various coupling and masses in terms of the four
parameters $G_N$, $G_V$, $m$ and $g$.

\subsection{ Kinetic terms and dimensions of the physical fields in the gravity sector }

  The Lagrangian of the gravity sector in Eqn.(5.\ref{LAGRAV})
  contains the vierbeins $e^\mu_a$ that define the physical metric
  in the form

\refstepcounter{equation}
$$
L_r = {1 \over 16 \pi G_N} r, \eqno{(\thesection.\theequation)}
\label{7.1}
$$
where $r$ is the scalar curvature in the conventional Riemannian
geometry.

  We next consider the kinetic terms of the tensor field $v^{\mu \nu}$
($s^{\mu \nu}$) by collecting together terms quadratic in these
fields and quadratic in their derivatives. They lead to the
partial Lagrangian

\refstepcounter{equation}
$$
\begin{array}{ccc}
L_v &~=~& - ( 4 \pi G_N)^{-1} \nabla_\lambda v_{\mu \nu}
\nabla^\lambda v^{\mu \nu} +  (2 \pi G_N)^{-1} \nabla_\lambda
v_{\mu \nu} \nabla^\mu v^{\lambda \nu} \\
&& + (4 \pi G_N)^{-1}
\partial_\mu v^\nu_\nu \partial^\mu v^\rho_\rho
-(2\pi G_N)^{-1} \partial_\mu v^\rho_\rho \nabla_\nu v^{\mu \nu} \\
&& -( 4\phi G_N)^{-1} m^2 ( v^{\mu \nu} v_{\mu \nu} - v^\mu_\mu
v^\nu_\nu ).
\end{array}
\eqno{(\thesection.\theequation)} \label{7.2}
$$

To obtain the correct kinetic coefficient for $v^{\mu \nu}$, we
need to define
\refstepcounter{equation}
$$
v^{\mu \nu} \longrightarrow \sqrt { \pi G_N} v^{\mu \nu}.
\eqno{(\thesection.\theequation)} \label{7.3}
$$
With this redefinition, which incidently, gives the correct
dimension to $v^{\mu \nu}$ ( because of $G_N$) and yields the
Fierz-Pauli Lagrangian for a spin-2 tensor field of mass $m$ in
curved spacetime.

The kinetic terms for the scalar fields are to be found from the
Eqn.(A.4.6) in the Appdendix A.4. The field $\phi_+$ needs special
treatment as in the conventional Kaluza-Klein theory, since this
field tends to 1 in vacuum. The following field redefinitions will
give the fields correct physical dimensions,

\refstepcounter{equation}
$$
\begin{array}{ccc}
\phi_+ &\longrightarrow& \exp{(2 G_V \sigma)} \\
\phi_- &\longrightarrow& 2  G_V  \phi_-
\end{array}
\eqno{(\thesection.\theequation)} \label{7.4}
$$

The kinetic part of the scalar Lagrangian then becomes

\refstepcounter{equation}
$$
\begin{array}{ccc}
{\cal L}_\phi & = & - {1 \over 2} ( \exp{( 4  G_V  \sigma)} + 4
G^2_V \phi^2_-)^{-1}
( \exp {(2  G_V  \sigma)} \partial_\mu \sigma \partial^\mu \sigma \\
&~& + {1 \over 2} \partial_\mu \phi_- \partial^\mu \phi_-) - m^2
(G_V )^{-1} ( 1+ 2G_V \phi^2_- \exp{(- 2G_V \sigma}))^{-2},
\end{array}
\eqno{(\thesection.\theequation)} \label{LAPHI}
$$
By expanding the non-linear factors in power of $\sigma $ and
$\phi_-^2$ and retaining the lowest order terms, we obtain

\refstepcounter{equation}
$$
\begin{array}{ccc}
{\cal L}_\phi & = & - {1 \over 2} \partial_\mu \sigma \partial^\mu
\sigma  -  {1 \over 2} \partial_\mu \phi_- \partial^\mu \phi_-  -
m^2 (G_V)^{-1}\phi_-^2 - m^2(G_V)^{-1} ...,
\end{array}
\eqno{(\thesection.\theequation)} \label{LAPHI2}
$$
which represents the kinetic part of the massless Brans-Dicke
scalar field $\sigma$ and the massive $\phi_-$. It is to be noted
that it also contains a cosmological constant term $m^2/G_V$.

  Next, we consider the kinetic and mass terms for the vector
fields  $a_{\mu \pm}$ to be found in Eqn.(A.4.\ref{A.4.4})
\refstepcounter{equation}
$$
{\cal L}_a  =  - {1 \over 4 G_V^2 } ( a^{\mu\nu}_+ a_{\mu \nu +} + a^{\mu \nu}_- a_{\mu \nu -} \\
 + 2 m^2 a^\mu_- a_{\mu -} ) \eqno{(\thesection.\theequation)}
\label{7.6}
$$

With the field redefinition
\refstepcounter{equation}
$$
a_{\mu \pm} \longrightarrow G_V a_{\mu \pm}
\eqno{(\thesection.\theequation)} \label{7.7}
$$
yields the massless vector field $a_{\mu +}$ and massive vector
field $a_{\mu -}$ of mass $m/\sqrt{2}$.

   The constant with dimension of mass $G_V$ gives the scalar and
vector fields correct physical dimensions in exactly the same way
as the Newton gravitational $G_N$ endows the correct dimension the
massive tensor field $ v^{\mu \nu}$.

\subsection { Field redefinition in the gauge sector and gauge coupling }

   The vector and scalar fields in the gauge sector have the correct
dimensions. However, those fields must also be redefined to have
the standard kinetic terms:

\refstepcounter{equation}
$$
\begin{array}{ccc}
   b_{\mu \pm} &\longrightarrow & g b_{\mu \pm}\\
   \eta        &\longrightarrow & g \eta \\
   \bar \eta   &\longrightarrow & g \bar \eta
\end{array}
 \eqno{(\thesection.\theequation)} \label{7.8}
$$

  The partial gauge Lagrangian that contains only gauge vector and
scalar fields comes from ${\cal L}_{G1}$, ${\cal L}_{G2}$, ${\cal
L}_{G3}$ in the Appendix A.5. After redefinitions, it is given by

\refstepcounter{equation}
$$
\begin{array}{ccc}
   {\cal L}_{b \eta} &= & - {1 \over 4} b_{\mu \nu +} b^{\mu\nu}_+ - {1 \over 4}  b_{\mu \nu -} b^{\mu \nu}_-  \\
   &~& - ( \phi^2_+ + \phi_-^2)^{-1}({\cal D}^\mu \bar \eta)({\cal D}_\mu \eta)
- {g^2 \over 2}(\phi_+^2 + \phi_-^2) ( \bar \eta \eta - (m/g)^2)^2
\end{array}
 \eqno{(\thesection.\theequation)} \label{7.9}
$$
Implementing spontaneous symmetry breaking in the standard
fashion, we find the mass of the surviving Higgs scalar to be $
\sqrt{2} m $. The mass of $b_{\mu-}$ equals $2m$ and $b_{\mu +}$
remains massless.

From the fermionic Lagrangian ( 6.\ref{A.6.11}) we find that the
vector gauge fields $b_{\mu\pm}$ couple to the vector and axial
vector currents $j^\mu_{\pm}$ and $ j^\mu_{5\pm} $, where

\refstepcounter{equation}
$$
\begin{array}{ccc}
   j^\mu_+ &= & - g\bar \psi \gamma^\mu \psi \\
   j^\mu_{5+} &= & -i g\sqrt{ \pi G_N} \bar \psi \gamma^\nu \gamma^5 v^\mu_\nu  \psi \\
   j^\mu_- &= & g\sqrt{ \pi G_N} \bar \psi \gamma^\nu v^\mu_\nu \psi \\
   j^\mu_{5-} &= & -ig \bar \psi \gamma^\mu\gamma^5 \psi
\end{array}
 \eqno{(\thesection.\theequation)} \label{7.10}
$$

   When the massive gravity field vanishes, as expected, the
vector gauge field $b_{\mu +}$ couples only to the vector current
$j^\mu_+$, the vector gauge field $b_{\mu-}$ couples only to the
axial current $j^\mu_{5-}$.

  The massive gravity field contributes a axial vector part to the
current coupled to the $b_{\mu +}$ and a vector part to the
current coupled to the $b_{\mu -}$. The magnitude of these
contributions is determined by the Newton gravitational constant

\subsection{ Parity violating interactions due to extended gravity}

  The vector and axial vector currents of the fermionic fields
  coupled to the vector fields $a_{\mu \pm}$ of the gravity sector are
  given respectively by
\refstepcounter{equation}
$$
\begin{array}{ccc}
   J^\mu_+ &= & - m \pi G_N G_V \bar \psi \gamma^a Y^{[\nu, \mu]}_{ab} s^b_\nu \psi
    - {1\over 2} \sqrt{2} g G_V  \bar \psi \gamma^\mu ( \bar \eta + \eta) \psi \\
           & ~ & - {i \over 2} \sqrt{2\pi G_N} G_V  \bar \psi
           \gamma^\nu v^\mu_\nu ( \bar \eta - \eta) \psi
\end{array}
 \eqno{(\thesection.\theequation)} \label{7.11}
$$
and
\refstepcounter{equation}
$$
\begin{array}{ccc}
   J^\mu_{5+} &= & i m \pi G_N G_V \bar \psi \gamma^a \gamma^5 X^{[\nu, \mu]}_{ab} s^b_\nu \psi
           - {1\over 2} \sqrt{2} g G_V  \bar \psi \gamma^\mu
           \gamma^5 ( \bar \eta + \eta) \psi \\
           & ~ & + {i \over 2} \sqrt{2\pi G_N} G_V  \bar \psi
           \gamma^\nu \gamma^5 v^\mu_\nu ( \bar \eta - \eta) \psi
\end{array}
 \eqno{(\thesection.\theequation)} \label{7.12}
$$
The vector current coupled to $ a_{\mu -}$ is
\refstepcounter{equation}
$$
\begin{array}{ccc}
   J^\mu_- &= & - m \pi G_N G_V  \bar \psi \gamma^a X^{[\nu, \mu]}_{ab} s^b_\nu \psi
            + {i\over 2} \sqrt{2} g G_V \bar \psi \gamma^\mu
           ( \bar \eta - \eta) \psi \\
           & ~ & - {1 \over 2} \sqrt{2\pi G_N} G_V \bar \psi
           \gamma^\nu v^\mu_\nu ( \bar \eta + \eta) \psi
\end{array}
 \eqno{(\thesection.\theequation)} \label{7.13}
$$
The axial vector current coupled to $ a_{\mu -}$ is
\refstepcounter{equation}
$$
\begin{array}{ccc}
   J^\mu_{5-} &= & - i m \pi G_N G_V \bar \psi \gamma^a \gamma^5 Y^{[\nu, \mu]}_{ab} s^b_\nu \psi
    + {i\over 2} \sqrt{2} g G_V  \bar \psi \gamma^\mu ( \bar \eta -  \eta) \psi \\
           & ~ & + {1 \over 2} \sqrt{2\pi G_N} G_V \bar \psi
           \gamma^\nu v^\mu_\nu ( \bar \eta + \eta) \psi
\end{array}
 \eqno{(\thesection.\theequation)} \label{7.14}
$$

\section{ Summary and conclusions}
\setcounter{equation}{0}

The main part of this paper is the geometric formulation of a
two-sheeted space-time within the framework of Conne's NCG. To
start with, we have  reviewed algebraic formulation of the
conventional differential geometry in order to set the stage for
its generalization to noncommutative case.

    Our starting point for the two-sheeted space-time is the choice
    of the algebra ${\cal A} = C^\infty(C, {\cal M}) \bigoplus
    C^\infty(C, {\cal M})$ as part of a spectral triple. However,
our model leads us to a subalgebra where the two functions on two
sheets are complex conjugates of each other. We are led to this
subalgebra as the underlying mathematical structure of the
two-sheeted space-time dictated by physics. Our previous pure
gravity theories on two-sheeted space-time \cite{VW1,VW2, LVW}
were constructed from a subalgebra of ${\cal A}$, which consisted
of a pair of real functions. In the present paper, the
construction of a spontaneously broken gauge theory requires that
the fifth component of the gauge one-form must be a generalized
complex valued function, with the values on the two sheets being
complex conjugates of each other. To be mathematically consistent,
the same condition need to be imposed on the differential forms of
the theory. The algebra of the generalized function $F$ of the
form \refstepcounter{equation}
$$
\begin{array}{c}
      F
    \end{array} =
\left(\begin{array}{cc}
      f(x)  &   0   \\
      0   &  f^*(x)
    \end{array}\right) ,\hspace{.5cm}
 \eqno{(\thesection.\theequation)} \label{8.1}
$$
where $f(x)$ is a complex valued function, forms a closed
subalgebra of ${\cal A}$. Therefore, the noncommutative geometry
can be constructed consistently with this subalgebra to describe
the physics of chiral spinors and gauge fields coupled to gravity
on two-sheeted space-time. Remarkably, from the standpoint of
physics, the restriction to this subalgebra is also required by
the signature of the kinetic terms of the massive modes in the
theory. Thus, starting from a naturally general algebra, we have
ended up with a restriction to a subalgebra to have a physically
meaningful theory. Therefore, our results show that there is an
intimate interplay between the mathematically consistent structure
and physics in NCG.

   The present theory also requires the definition of the wedge
product to retain the Higgs quartic potential and to avoid the
"junk forms" at the same time. The theory also requires a
consistent involutive operator, that makes the theory consistent
and the scalar products to be diagonal. The present theory also
takes advantage of a minimal solution of the Cartan-Maurer
structure equations, where the basic geometric quantities such as
metric, connection, torsion and curvature are completely expressed
in terms of the generalized vielbeins without imposing any
constraint on its most general form. Physics requires that with
such solutions, the contribution from torsion must be included to
give kinetic terms to the physical fields. In other words, the
mathematical structure of torsion is required by physics in this
theory.

  It is worth noting here that, although the above structure is
rather unique in its simplicity and consistency, it is possible to
construct the theory in an alternative way. For example, if the
fully anti-symmetric wedge product is desired, one may resort to a
Kahler structure of the internal space consisting of a pair of
complex conjugates \cite{ZZbar,WIGNER}. In such a theory, the
involutive operation can be chosen consistently to be the
hermitian conjugate. This approach, in the context of the full
standard model, is currently under investigation.

From the point of view of physics, the model presented here
possesses a rich and complex structure that merits further
exploration in many ways. The generalized gravity sector that
includes massive tensor, vector and scalar fields, introduces
simultaneous vector and axial vector couplings to matter fields.
Consequently, we have the possibility of parity violating
interactions due to gravity. In the context of a full standard
model, we speculate that these interactions may provide the much
sought CP-violating interactions in the early universe. This is
further supported by the fact that such interactions involve $G_N$
as the coupling. Consequently, we expect them to be small and
hence correct order of magnitude. Incidentally, the idea of two
interacting gravitons, one massless and the other massive, goes
back to Isham, Salam and Strathdee \cite{Isham}. More recently,it
has been a subject of study from various theoretical points of
view\cite{Georgi, Chams}. It has also been pointed out by Damour
and Kogan that bigravity , which is the simplest form of
multigravity arises naturally in several different physical
contexts such as brane configurations, Kaluza-Klein reductions and
non-commutative geometric models (ours is an example),
\cite{Damour}.

In the context of cosmology, it is worth noting that the
gravitational sector comes with a scalar field, an associated
scalar potential and a cosmological constant term. A scalar field
and a scalar potential are invoked in current literature to
account for inflation and dark energy. Whether the potential in
the model presented here has the desired features or not needs
further exploration.

{\bf ACKNOWLEDGEMENTS}
 This work was supported in part (KCW) by the U.S.Department of Energy (DOE)under
 contract no. DE-FG02-85ER40237. Preliminary part of this work was
 done in collaboration with James A. Javor (James A. Javor and Kameshwar C. Wali,
 Construction of Action Functionals in Non-Commutative Lorenntzian Geometry, SU4240-692,
 unpublished). One of the authors (KCW) would like to thank A. Connes and J. Madore for
 many helpful discussions during the course of this work. He is
 also indebted to Joel Rozowsky for discussions and his help with the manuscript.

\newpage

 \underline{{\bf APPENDIX}}

\vskip .3cm
    We collect together in this appendix some formulae used in our
    calculations in the main text. We also present various Lagrangians needed to construct
    the action functionals.The field redefinitions (5.\ref{FIREDEF1})
    and (6.\ref{FIREDEF2}) are taken into account to obtain the correct
    signatures for the kinetic terms in the final actions.

\vskip .5cm

 {\bf {A.1 Metric components $G^{\mu \nu}$ and
$G_{\mu \nu }$ }}

    Expressed in terms of the physical variables defined in
Sect.5.2, the even and odd components of the metric are given by
$$
 G^{\mu\nu} = G^{\mu\nu}_+ {\bf 1} + G^{\mu\nu }_- {\bf r}~~,~~
 G_{\mu \nu} = G_{\mu \nu +} {\bf 1} + G_{\mu \nu -} {\bf r },
\eqno{(A.1.1)} \label{A.1.1}
$$
where
$$
G^{\mu\nu}_+ = g^{\mu \nu} - g^{\sigma\lambda} v^\mu_\sigma
v^\nu_\lambda  ~~,~~ G^{\mu \nu}_- = i (g^{\mu \sigma}
v_\sigma^\nu + g^{\nu \sigma} v^\mu_\sigma ), \eqno{(A.1.2)}
\label{A.1.2}
$$
$$
G_{\mu\nu+} =  g_{\mu \nu} + (e^a_\mu \eta_{ab} s^b_\lambda )
v^\lambda_\nu + ( e^a_\nu \eta_{ab} s^b_\lambda ) v^\lambda_\mu +
( v^\lambda_\mu v^\sigma_\nu - \delta^\lambda_\mu
\delta^\sigma_\nu ) s^a_\lambda \eta_{ab} s^b_\sigma  ,
 \eqno{(A.1.3)}
 \label{A.1.3}
$$
$$
G_{\mu\nu-} = i ( e^a_\mu \eta_{ab} s^b_\nu + e^a_\nu \eta_{ab}
s^b_\mu + v^\lambda_\mu s^a_\lambda \eta_{ab} s^b_\nu +
v^\lambda_\nu s^a_\lambda \eta_{ab} s^b_\mu ).
 \eqno{(A.1.4)}
 \label{A.1.4}
$$
It is to be noted that $ G^{\mu \nu}, G_{\mu \nu} $ are symmetric,
$$
G^{\mu \nu}_{\pm} = G^{\nu \mu}_{\pm} ~~,~~ G_{\mu\nu\pm} =
G_{\nu\mu \pm}.
 \eqno{(A.1.5)}
 \label{A.1.5}
$$

{\bf {A.2 Components of $D\Gamma^A $ }}
From
$$
     \Gamma^A = \Gamma^M E_M^A ~~,
     \eqno{(A.2.1)}
     \label{A.2.1}
$$
Eqn.(4.\ref{GENDEVCOMP}) gives
$$
\begin{array}{ccc}
(D\Gamma^A)_{\mu\nu} & = & {1 \over 2} ( D_\mu E^A_\nu - D_\nu
E^A_\mu ), \\
(D\Gamma^A)_{\mu 5} & = & {1 \over 2} ( D_\mu E^A_5 - D_5
E^A_\mu) = - (D\Gamma^A)_{5\mu} \\
(D\Gamma^A)_{55} & = & m( E^A_5 + {\tilde E}^A_5 ).
\end{array}
 \eqno{(A.2.2)}
 \label{A.2.2}
$$
Transforming Eqn.(A.2.2) into the locally flat frame, one obtains
the components $(D\Gamma^A)_{BC}$ as follows

$$
\begin{array}{ccc}
(D\Gamma^a)_{bc} & = &  {1 \over 2} E^\mu_b E^\nu_c [
( D_\mu E^a_\nu - D_\nu E^a_\mu) + ( A_\nu D_5E^a_\mu - A_\mu D_5 E^a_\nu)],   \\
(D\Gamma^a)_{b {\dot 5}} & = & -{ 1 \over 4} (D_5E^a_\mu)
E^5_{{\dot 5}} ( {\tilde E}^\mu_b + E^\mu_b )
= - (D\Gamma^a)_{{\dot 5} b}  , \\
(D\Gamma^a)_{{\dot 5}{\dot 5}} & = & 0       , \\
(D\Gamma^{{\dot 5}})_{bc} &  = & {1 \over 2} E^\mu_b E^\nu_c [ (
D_\mu A_\nu - D_\nu A_\mu) \Phi + (A_\nu D_5 ( A_\mu \Phi) - A_\mu
D_5(A_\nu\Phi))], \\
 (D\Gamma^{{\dot 5}})_{b{\dot 5}} &= & { 1 \over 2 } E^5_{{\dot
5}} [{ 1\over 2} ({\tilde E}^\mu_b + E^\mu_b) ( D_\mu E^{{\dot
5}}_5 - D_5 E^{{\dot 5}}_\mu) + m  ({\tilde E}^5_b - E^5_b)
(E^{{\dot 5}}_5 + {\tilde E}^{{\dot 5}}_5)],\\
(D\Gamma^{{\dot 5}})_{{\dot 5}{\dot 5}} & = & m {\tilde
E}^5_{{\dot 5}} E^5_{{\dot 5}}( E^{{\dot 5}}_5 + {\tilde E}^{{\dot
5}}_5).
\end{array}
\eqno{(A.2.3)}\label{A.2.3}
$$

We note that, since the lower indexed $\Gamma_A$ matrix is defined
as
$$
\Gamma_A = \eta_{AD}\Gamma^D,
$$

we use the following notation

$$
(D\Gamma_A)_{BC} = \eta_{AD} (D\Gamma^D)_{BC}.
 \eqno{(A.2.4)}
 \label{A.2.4}
$$

Using the definitions of the generalized vielbeins in
Eqns.(4.\ref{VIELBEIN}) and (4.\ref{VIELBEIN-INV}) and expressing
them in terms of the new variables with the field redefinitions
Eqn.(5.\ref{FIREDEF1}) and (6.\ref{FIREDEF2}), we find the even
and odd components of $(D\Gamma^A)_{BC}$. They are

$$
\begin{array}{ccc}
(D\Gamma_a)_{bc+}&=&{1 \over 2}[X^{[\mu,\nu]}_{bc} P_{a\mu\nu}

- Y^{[\mu, \nu]}_{bc}Q_{a\mu\nu} ]\\
(D\Gamma_a)_{bc-}&=&  {i \over 2} [X^{[\mu, \nu]}_{bc} Q_{a\mu
\nu} + Y_{bc}^{[\mu, \nu]} P_{a\mu\nu}],
\end{array}
\eqno{(A.2.5)}
 \label{ADGAM0}
$$

$$
\begin{array}{ccccc}
 (D\Gamma_a)_{b {\dot 5}+}  &~~=~~& - m \phi_- e^\mu_b s_{a\mu} (
\phi^2_+ + \phi_-^2 )^{-1} ~~=~~-(D\Gamma_a)_{{\dot 5} b+}, \\

(D\Gamma_a)_{b{\dot 5}-} &~~=~~& -  i m \phi_+ e^\mu_b s_{a\mu} (
\phi^2_+ + \phi^2_- )^{-1} ~~=~~-(D\Gamma_a)_{{\dot 5}b -},
\end{array}
 \eqno{(A.2.6)}
 \label{A.2.6}
$$
$$
(D\Gamma^a)_{{\dot 5}{\dot 5}\pm} ~~=~~0 \eqno{(A.2.7)}
\label{A.2.7}
$$
$$
\begin{array}{ccc}
(D\Gamma_{{\dot 5}})_{bc+} &~~=~~& {1 \over 2}(
X^{[\mu,\nu]}_{bc}(a_{\mu\nu+} \phi_+ - a_{\mu\nu-} \phi_-) -
 Y^{[\mu, \nu]}_{bc} (a_{\mu\nu+} \phi_- + a_{\mu\nu-} \phi_+))\\
 &~~~~~&
- m(  X^{[\mu, \nu]}_{bc} a_{\nu-} a_{\mu+}\phi_- +  Y^{[\mu,
\nu]}_{bc} a_{\nu+} a_{\mu-}\phi_+) \\
(D\Gamma_{{\dot 5}})_{bc-} &~~=~~& {i \over 2}[(
X^{[\mu,\nu]}_{bc}(a_{\mu\nu+} \phi_- + a_{\mu\nu-} \phi_+) +
Y^{[\mu, \nu]}_{bc} (a_{\mu\nu+} \phi_+ - a_{\mu\nu-} \phi_-))\\
 &~~~~~& + 2 m(  X^{[\mu, \nu]}_{bc} a_{\nu+} a_{\mu-}\phi_+ -  Y^{[\mu,
\nu]}_{bc} a_{\nu-} a_{\mu+}\phi_-)],
\end{array}
\eqno{(A.2.8)} \label{A.2.8}
$$
$$
\begin{array}{ccc}
(D\Gamma_{{\dot 5}})_{b{\dot 5}+}&=&( \phi^2_+ +
\phi^2_-)^{-1}[ {1 \over 2} (\phi_+ \partial_\mu \phi_+ +
\phi_- \partial_\mu \phi_-) e^\mu_b \\
&~& - m  e^\mu_b \phi_-( a_{\mu-}\phi_+ + a_{\mu+}\phi_-) + 2m
\phi_- \phi_+( a_{\mu-} e^\mu_b + a_{\mu +} v^\mu_b)] \\
(D\Gamma^{{\dot 5}})_{b{\dot 5}-}&=& i( \phi^2_+ + \phi^2_-)^{-1}[
{1 \over 2} (\phi_+ \partial_\mu \phi_-
- \phi_- \partial_\mu \phi_+) e^\mu_b \\
&~& -  m  e^\mu_b \phi_+( a_{\mu-}\phi_+ + a_{\mu+}\phi_-) + 2m
\phi_+^2(  a_{\mu-} e^\mu_b + a_{\mu +} v^\mu_b)],
\end{array}
 \eqno{(A.2.9)}
 \label{A.2.9}
$$

$$
(D\Gamma_{{\dot 5}})_{{\dot 5}{\dot 5}+} ~~=~~ 2m \phi_+ (\phi_+^2
+ \phi^2_- )^{-1} ~~,~~ ( D\Gamma^{{\dot 5}})_{{\dot 5}{\dot 5}- }
= 0, \eqno{(A.2.10)}
 \label{ADGAME}
$$
where we have defined
$$
a_{\mu \nu \pm} = {1 \over 2} (\partial_\mu a_{\nu\pm} -
\partial_\nu a_ {\mu \pm}),
 \eqno{(A.2.12)}
 \label{A.2.12}
$$
and $P_{a\mu\nu}$, $ Q_{a\mu\nu}$, $X^{[\mu, \nu]}_{ab}$,
$Y^{[\mu, \nu]}_{ab}$ are defined in Eqn.(5.\ref{5.13}).

    The following formulae for $X^{[\mu, \nu]}_{bc}$ and $Y^{[\mu,
\nu]}_bc$ will be useful in the torsion and gauge Lagrangian
calculations
$$
\begin{array}{ccc}
X_{bc}^{[\mu, \nu]} X^{bc[\rho, \tau]} &=& 2( g^{\mu \rho} g^{\nu
\tau} - g^{\mu \tau} g^{\nu \rho} + 2  v^{\mu \tau} v^{\nu \rho}
- 2 v^{\mu\rho} v^{\nu\tau} \\
&+& <v^\mu, v^\rho><v^\nu, v^\tau> - <v^\mu, v^\tau><v^\nu,
v^\rho>)
\end{array}
\eqno{(A.2.13)} \label{A.2.13}
$$
$$
\begin{array}{ccc}
X_{bc}^{[\mu, \nu]} Y^{bc[\rho, \tau]} &=& 2 ( g^{\mu \rho}
v^{\nu\tau} - g^{\mu\tau}v^{\nu\rho} + v^{\mu\rho} g^{\nu \tau} -
v^{\mu \tau} g^{\nu \rho} + v^{\nu \rho}<v^\mu, v^\tau>
\\ &~& - v^{\nu\tau}<v^\mu, v^\rho> + v^{\mu \tau}<v^\nu, v^\rho>
- v^{\mu\rho}<v^\nu, v^\tau>),
\end{array}
\eqno{(A.2.14)} \label{A.2.14}
$$
$$
\begin{array}{ccc}
Y_{bc}^{[\mu, \nu]} Y^{bc[\rho, \tau]} &=&  2(g^{\mu \rho}<v^\nu,
v^\tau> - g^{\mu\tau}<v^\nu, v^\rho> + g^{\nu \tau}<v^\mu, v^\rho>
\\ &~&
- g^{\nu\rho}<v^\mu, v^\tau> + 2v^{\mu \rho}v^{\nu\tau} -
2v^{\mu\tau} v^{\nu\rho})
\end{array}
\eqno{(A.2.15)} \label{A.2.15}
$$
   Using the formulae (A.2.\ref{ADGAM0})-(A.2.\ref{ADGAME}), it is straightforward to
calculate all components of the connection, torsion and scalar
curvature. \vskip 1cm

{\bf { A.3 Curvature Lagrangian }}

    For the calculation of the scalar curvature in Eqn.(5.\ref{SCURV}) the
components $\Omega_{abc}$ and $\Omega_{a {\dot 5} b}$ of the
connection $\Omega_{AB}$ are necessary. The explicit expressions
of the projections of these components are given by
$$
\begin{array}{ccc}
\Omega_{a {\dot 5}b + }& = & - 2 (D\Gamma_a)_{{\dot 5} b +} = -2 m
\phi_- e^\mu_b s_{a\mu} ( \phi^2_+ + \phi^2_- )^{-1} \\
\Omega_{a {\dot 5}b - }& = &  2 (D\Gamma_a)_{{\dot 5} b -} = 2 i m
\phi_+ e^\mu_b s_{a\mu} ( \phi^2_+ + \phi^2_- )^{-1}
\end{array}
\eqno{(A.3.1)}\label{ACON0}
$$
$$
\begin{array}{ccc}
\Omega_{abc + }& = & -((D\Gamma_a)_{bc+} - (a \leftrightarrow b) +
(a \leftrightarrow c))~~~~~~~~~~~~~~~~~~~~~~~~~~~~~~ \\
& = & - {1 \over 2}( (  X^{[\mu, \nu]}_{bc} P_{a\mu\nu}
 -  Y^{[\mu, \nu]}_{bc} Q_{a\mu\nu}) - (a \leftrightarrow b) + (a \leftrightarrow c))\\
\Omega_{abc - }& = & -((D\Gamma_a)_{bc-} - (a \leftrightarrow b) +
(a \leftrightarrow c))~~~~~~~~~~~~~~~~~~~~~~~~~~~~~~\\
 & = & - {i \over 2} (( X^{[\mu, \nu]}_{bc} Q_{a\mu \nu}
+Y^{[\mu, \nu]}_{bc} P_{a\mu \nu}) - (a \leftrightarrow b) + (a
\leftrightarrow c)) ,
\end{array}
\eqno{(A.3.2)}\label{A.3.2}
$$
where $(a \leftrightarrow b)$ and $(a \leftrightarrow b)$ are
permutations of the indices of the given tensors.

  Now we can calculate the following terms
$$
\begin{array}{ccc}
(D\Omega_{ab})_{cd+} \eta^{ad} \eta^{bc} &=& - X^{ab[\mu, \nu]}(
{1 \over 2} \partial_\mu \Omega_{ab\nu +} + i m a_{\nu-}
\Omega_{ab
\mu - })\\
&~& - i Y^{ab [\mu, \nu]} ( {1 \over 2} \partial_\mu \Omega_{ab
\nu-} + im a_{\nu -} \Omega_{ab \mu +}),
\end{array}
 \eqno{(A.3.3)} \label{A.3.3}
$$
where $\Omega_{ab\mu \pm}$ can be expressed in terms of
$\Omega_{abc\pm}$ as follows
$$
\begin{array}{ccc}
\Omega_{ab\mu +} & = & ( e^c_\mu + e^c_\nu s^{\nu \rho} v_{\rho
\mu}) \Omega_{abc+} + i s^c_\mu \Omega_{abc-} \\
\Omega_{ab\mu -} & = & ( e^c_\mu + e^c_\nu s^{\nu \rho} v_{\rho
\mu}) \Omega_{abc-} + i s^c_\mu \Omega_{abc+}
\end{array}
\eqno{(A.3.4)} \label{ACONE}
$$
 Inserting $\Omega_{abc\pm}$ from Eqn.(A.3.\ref{A.3.2}) into Eqn.(A.3.\ref{ACONE}), we
 calculate $\Omega_{ab\mu\pm}$ and then $ (D\Omega_{ab})_{cd+} \eta^{ad}
 \eta^{bc}$, giving the following final expression

$$
{\cal L}_{R1} = ( 16 \pi G_N)^{-1} (D\Omega_{ab})_{cd+} \eta^{ad}
\eta^{bc}
$$
$$
\begin{array}{ccc}
{\cal L}_{R1} &=&  (64 \pi G_N)^{-1} ( e_{d\tau} + s_{d\lambda}
v^\lambda_\tau)
 [ (2 X^{ab[\rho, \tau]} \eta^{cd} - X^{bc[\rho,\tau]}
\eta^{ad}) \partial_\rho(X_{bc}^{[\mu, \nu]} P_{a\mu\nu}
- Y^{[\mu, \nu]} Q_{a\mu\nu})\\
&~& - ( 2 Y^{ab[\rho,\tau]} \eta^{cd} - Y^{bc[\rho, \tau]}
\eta^{ad}) \partial_\rho ( X^{[\mu, \nu]}_{bc} Q_{a\mu\nu} +
Y^{[\mu, \nu]}_{bc} P_{a\mu\nu})]\\
&~& +( 64 \pi G_N)^{-1}( X^{bc[\rho, \tau]} X_{bc}^{[\mu, \nu]}
\eta^{ad} + 2X^{ab[\rho, \tau]} X^{d~[\mu,\nu]}_{~b})(
-P_{a\mu\nu} \partial_\rho(e_{d\tau} + s_{d\lambda} v^\lambda_\tau)\\
&~& + Q_{a\mu\nu} \partial_\rho s_{d\tau}) + ( 64 \pi G_N)^{-1}(
X^{bc[\rho, \tau]} Y_{bc}^{[\mu, \nu]} \eta^{ad} + 2X^{ab[\rho,
\tau]} Y^{d~[\mu,\nu]}_{~b}\\
&~& + Y^{bc[\rho, \tau]} X_{bc}^{[\mu, \nu]} \eta^{ad} +
2Y^{ab[\rho, \tau]} X^{d~[\mu,\nu]}_{~b})
(P_{a\mu\nu}\partial_\rho s_{d\tau} +
Q_{a\mu\nu}\partial_\rho(e_{d\tau} + s_{d\lambda} v^\lambda_\tau))\\
&~& (64 \pi G_N)^{-1}( Y^{bc[\rho, \tau]} Y_{bc}^{[\mu, \nu]}
\eta^{ad}
+ 2Y^{ab[\rho, \tau]} Y^{d~[\mu,\nu]}_{~b}) \\
&~&(P_{a\mu\nu} \partial_\rho(e_{d\tau} + s_{d\lambda}
v^\lambda_\tau)
+ Q_{a\mu\nu} \partial_\rho s_{d\tau})\\
&~& + m ( 32 \pi G_N)^{-1} a_{\tau -} [(( X^{bc[\rho, \tau]}
X_{bc}^{[\mu, \nu]} - Y^{bc[\rho, \tau]} Y_{bc}^{[\mu,
\nu]})\eta^{ad} - 2 ( X^{ab[\rho, \tau]} X^{d~[\mu, \nu]}_{~b}\\
&~& - Y^{ab[\rho, \tau]} X^{d~[\mu, \nu]}_{-b}))( P_{a\mu\nu}
s_{d\rho}
+ Q_{a\mu\nu}(e_{d\rho} + s_{d\lambda} v^\lambda_\rho)) \\
&~& + (( Y^{bc[\rho, \tau]} X_{bc}^{[\mu, \nu]} - X^{bc[\rho,
\tau]} Y_{bc}^{[\mu, \nu]})\eta^{ad} + 2 ( Y^{ab[\rho, \tau]}
X^{d~[\mu, \nu]}_{~b} \\
&~&+ X^{ab[\rho, \tau]} X^{d~[\mu, \nu]}_{~b})) (
P_{a\mu\nu}(e_{d\rho} + s_{d\lambda} v^\lambda_\rho)  +
Q_{a\mu\nu} s_{d\rho})].
\end{array}
 \eqno{(A.3.5)} \label{A.3.5}
$$
$$
\begin{array}{ccc}
{\cal L}_{R2} &=& ( 16 \pi G_N)^{-1} ( (D\Gamma_a)_{ec+}
(D\Gamma_b)_{fd+} +
(D\Gamma_a)_{ec-} (D\Gamma_b)_{fd-}) \\
&~&(  \eta^{ab} \eta^{cd} \eta^{ef} + 3 \eta^{ad} \eta^{be}
\eta^{cf} + \eta^{ac} \eta^{bd} \eta^{ef})\\
&=& (64 \pi G_N )^{-1} (\eta^{ab}
X^{cd[\mu,\nu]}X^{[\rho,\tau]}_{cd} - 3 X^{bc[\mu, \nu]}
X^{a~[\rho,\tau]}_{~c} +
X^{ac[\mu,\nu]}X^{b~[\rho, \tau]}_{~c})\\
&~&[(P_{a\mu\nu} P_{b\rho\tau} - Q_{a\mu\nu} Q_{b\rho\tau})] \\
&~& + ( 64 \pi G_N )^{-1} (\eta^{ab}
Y^{cd[\mu,\nu]}Y^{[\rho,\tau]}_{cd} - 3 Y^{bc[\mu, \nu]}
Y^{a~[\rho,\tau]}_{~b} +
Y^{ac[\mu,\nu]}Y^{b~[\rho, \tau]}_{~c})\\
&~& [ - P_{a\mu \nu} P_{b\rho\tau} + Q_{a\mu \nu} Q_{b\rho\tau}]\\
&~&+ ( 64 \pi G_N)^{-1} ( -\eta^{ab} ( X^{cd [\mu, \nu]}
Y_{cd}^{[\rho, \tau]} - X^{cd [\rho, \tau]} Y^{[\mu, \nu]}_{cd}) +
2( X^{bc[\mu,
\nu]} Y^{a~[\rho,\tau]}_{~c}\\
&~& - X^{bc[\rho, \tau]} Y^{a~[\mu, \nu]}_{~c}))[ P_{a\mu\nu}
Q_{b\rho\tau} - Q_{a\mu \nu} P_{b\rho\tau}],
\end{array}
\eqno{(A.3.6)} \label{A.3.6}
$$
$$
\begin{array}{ccc}
 {\cal L}_{R3} &=& (8 \pi G_N)^{-1} [(D\Omega_{a{\dot 5}})_{{\dot 5}d+}\eta^{ad}
 + 4 ((D\Gamma_a)_{{\dot 5}d+} (D\Gamma_b)_{{\dot 5}
c+}\\
&~& + (D\Gamma_a)_{{\dot 5}d-} (D\Gamma_b)_{{\dot 5}
c-})(\eta^{ad} \eta^{bc} - \eta^{ac} \eta^{bd})] \\
&  = & - m^2 ( 4\pi G_N)^{-1}(\phi_+^2+\phi_-^2)^{-2}[
(\phi^2_+-\phi^2_-)
  ( s^{\mu\nu} s_{\mu\nu} -  s^\mu_\mu s^\nu_\nu )  \\
 &~& +\phi_+(\phi_+(v^{\mu\nu}s_{\mu\rho}v^{\rho\tau}s_{\tau\nu}
 - v^{\mu \nu} s_{\mu \nu}) + \phi_- s^{\mu\nu}s^\rho_\mu
 s_{\nu\rho})].
\end{array}
\eqno{(A.3.7)}
$$

{\bf { A.4 Torsion Lagrangian}}

    The contributions of the torsion to the total Lagrangian arise from
    the components $ T_{{\dot 5} bc} $ , $T_{{\dot 5} {\dot 5} b}$
and $ T_{{\dot 5}{\dot 5} {\dot 5}}$.

    With the physical reasoning that the gravity tensor $v_{\mu \nu} $
represents a field of spin 2, we  assume that the gravity tensor
field $v_{\mu \nu}$ is symmetric in the indices $\mu, \nu$, i.e.
$v_{\mu \nu} = v_{\nu \mu}$. This also implies $s_{\mu \nu} =
s_{\nu \mu}$ and hence,

$$
(D\Gamma_b)_{{\dot 5}c \pm} - (D\Gamma_c)_{{\dot 5} b \pm} = 0 ,
\eqno{(A.4.1)} \label{A.4.1}
$$

    Hence, the projections of the torsion components can be expressed in terms
of $D\Gamma$ as follows

$$
\begin{array}{ccc}
(T_{{\dot 5}bc})_\pm &=& (D\Gamma_{{\dot 5}})_{bc\pm}, \\
(T_{{\dot5}b {\dot 5}})_\pm &=& (D\Gamma_{{\dot 5}})_{{\dot 5}b \pm}, \\
(T_{{\dot 5}{\dot 5}{\dot 5}})_\pm &=& (D\Gamma_{{\dot 5}})_{{\dot
5}{\dot 5} \pm},
\end{array}
\eqno{(A.4.2)} \label{A.4.2}
$$

     Using the expressions of $D\Gamma$ in Appendix A.2, we obtain

$$
{\cal L}_{T1} ~=~ ( 4 G^2_V)^{-1}[(T_{{\dot 5}bc})_+ (T^{{\dot 5}
bc})_+ - (T_{{\dot 5}bc})_- (T^{{\dot 5}
bc})_-]~,~~~~~~~~~~~~~~~~~~~~~~~~~~
 \eqno{(A.4.3}) \label{A.4.3}
$$
$$
\begin{array}{ccc}
{\cal L}_{T1} &=& - (4 G^2_V)^{-1}( X_{bc}^{[\mu, \nu]}
X^{bc[\rho, \tau]} + Y_{bc}^{[\mu, \nu]} Y^{bc[\rho, \tau]})
[(\phi_+^2 +\phi^2_-)( {1 \over 4} a_{\mu\nu+} a_{\rho \tau_+} +
{1 \over 4}
a_{\mu \nu -} a_{\rho \tau -}\\
&~& + m^2 a_{\mu_-} a_{\nu+} a_{\rho-} a_{\tau +}) + m a_{\rho -}
a_{\tau +} ( 2 a_{\mu \nu +} \phi_+ \phi_- + a_{\mu \nu-} (
\phi_+^2 - \phi^2_-))]\\
&~& + m ( G^2_V)^{-1} X_{bc}^{[\mu, \nu]} Y^{bc[\rho, \tau]}[
(\phi_+^2 - \phi_-^2) (a_{\mu \nu+} a_{\tau -} a_{\rho +}  +
a_{\nu +} a_{\mu -} a_{\rho \tau +} )\\
&~& + 2 \phi_+\phi_-( a_{\mu \nu-} a_{\tau +} a_{\rho -} + a_{\rho
\tau -} a_{\nu -} a_{\mu +})] ,
\end{array}
\eqno{(A.4.4)} \label{A.4.4}
$$

$$
{\cal L}_{T2} ~=~ ( 2 G^2_V)^{-1} [(T_{{\dot 5}{\dot 5}c})_+
(T^{{\dot 5} {\dot 5} c})_+ - (T_{{\dot 5}{\dot 5} c})_- (T^{{\dot
5} {\dot 5} c})_-], \eqno{(A.4.5)} \label{A.4.5}
$$
$$
\begin{array}{ccc}
{\cal L}_{T2} &=& - ( 2 G^2_V )^{-1}( \phi_+^2 + \phi_-^2)^{-1}[
{1 \over 4} g^{\mu \nu}(\partial_\mu \phi_+
\partial_\nu \phi_+ + \partial_\mu \phi_- \partial_\nu \phi_-) +
m^2( a^\mu_- a_{\mu -} (4 -3 \phi_+^2)\\
&~& + \phi^2_- a^\mu_+ a_{\mu+}  - 2 \phi_+ \phi_- a_{\mu-}
a^\mu_+ +  4 a_{\mu+} v^\mu_\rho( v^{\rho \lambda}a_{\lambda +} -
a^\rho_+\phi_+\phi_- + 2a^{\rho}_-
- a^\rho_- \phi_+^2))\\
&~&+ m ( a^\mu_- \phi_+ + 2 a_{\rho +} v^{\rho \mu} \phi_+ -
a^\mu_+ \phi_-)\partial_\mu \phi_-]
\end{array}
\eqno{(A.4.6)} \label{A.4.6}
$$

$$
{\cal L}_{T3} ~=~-( 4 G^2_V )^{-1} (T_{{\dot 5}{\dot 5}{\dot
5}})_+ T^{{\dot 5}{\dot 5}{\dot 5}}_+ ~=~ - {m^2 \over G^2_V k^2}
\phi_+^2(\phi_+^2 + \phi^2_-)^{-2}. \eqno{(A.4.7)} \label{A.4.7}
$$

{\bf { A.5 Gauge Lagrangian } }

   The Lagrangian terms in Eqn.(6.\ref{LAGAU2}) can be calculated
   as follows
$$
 {\cal L}_{G1}  =  - {1 \over 4 g^2} ( g^{\mu \rho} + g^{\sigma
 \lambda} v^\mu_\sigma v^\rho_\lambda) ( g^{\nu \tau} + g^{\sigma
 \lambda} v^\nu_\sigma v^\tau_\lambda)
  ( b_{\mu \nu+} b_{\rho \tau+} + b_{\mu \nu-} b_{\rho \tau
  -})
\eqno{(A.5.1)}
$$

$$
\begin{array}{ccc}
{\cal L}_{G2} & = & - { 1 \over g^2} ( \phi^2_+ + \phi^2_-)^{-1} (
g^{\mu \nu} {\cal D}_\mu \eta {\cal D}_\nu \bar \eta \\
&~& + 2i g^\mu \sigma v^\nu_\sigma b_{\mu +} ( (\bar \eta -m)
{\cal
D}_\mu \eta - (\eta -m) {\cal D}_\mu \bar \eta) \\
&~& + 4 g^{\sigma \tau} v^\mu_\sigma v^\nu_\tau b_{\mu +} b_{\nu
+} ( \eta -m ) ( \bar \eta -m))
\end{array}
\eqno{(A.5.2)}
$$

$$
{\cal L}_{G3} =  {1 \over 2g^2} ( \phi_+^2 + \phi_-)^{-1} ( \bar
\eta \eta - m^2)^2. \eqno{(A.5.3)}
$$

$$
\begin{array}{ccc}
{\cal L}_{G4} & = & { 1 \over g^2} ( X^{[\mu, \nu]}_{ab}
X^{ab[\rho,
\tau]} + Y^{[\mu, \nu]}_{ab} Y^{ab[\rho, \tau]})\\
&~& [ ( ( b_{\mu \nu +} - i b_{\mu \nu -})a_\rho ( {\cal D}_\tau
\eta + 2i (\eta -m) b_{\tau -}) + c.c) \\
&~& +  ({1\over 2} a_\mu a^*_\rho ({\cal D}_\nu \bar \eta \bar
({\cal D}_\tau \eta + 2i (\eta -m) b_{\tau -}) + 4 b_{\nu-} (\eta
-m)(\bar \eta -m)) + c.c)] \\
&~& - {1 \over 2g^2} (\phi^2_+ + \phi^2_-)^{-1} g^{\mu\nu}[( X_\mu
+ X^*_\mu) ( Y_\nu + Y^*_\nu) + ( X_\mu -  X^*_\mu + 4im
v^\rho_\mu b_{\rho +}) (
Z_\nu - Z^*_\nu)\\
&~& + {1 \over 2} ( Y_\mu + Y^*_\mu)(Y_\nu +  Y^*_\nu) - {1 \over
2} ( Z_\mu - Z^*_\mu)(Z_\nu - Z^*_\nu)]
\end{array}
\eqno{(A.5.4)}
$$
where
$$
\begin{array}{ccc}
a_\mu & = &   a_{\mu+} + i a_{\mu -} \hskip 2cm \\
X_\mu & = &  ( {\cal D}_\mu - 2i v^\nu_\mu b_{\nu +})\eta \\
Y_\mu & = &  ( a_\mu + i v^\nu_\mu a_\nu) (\eta -\bar
\eta)(\eta -m)\\
Z_\mu & = & ( a_\mu - i v^\nu_\mu a_\nu)(\eta + m)(\eta + \bar
\eta -2m ),
\end{array}
\eqno{(A.5.5)}
$$
$c.c$ denotes the complex conjugates.

\vskip 1.5cm

{\bf {A.6 Fermionic Lagrangian }}

The following identities will be useful for the calculations of
the Lagrangian in Eqn.(6.\ref{LAFE}).
$$
\begin{array}{ccc}
\bar \Psi\Gamma^a A \Psi &~=~& \bar \psi \gamma^a (A_+ + \gamma^5 A_-)\psi,\\
\bar \Psi \Gamma^{{\dot 5}}A \Psi &~=~& \bar \psi (\gamma^5 A_+ +
A_-)\psi,
\end{array}
\eqno{(A.6.1)} \label{A.6.1}
$$
where
$$
A = A_+ {\bf 1} + A_-{\bf r}.
$$

   To prove these identities we can take the trace over the $Z_2$
indices then express $\psi_L, \psi_R$ in terms of the 4-component
Dirac spinor and the parity projection operator $ {1 \over 2}(1
\pm \gamma^5)$.

To calculate the terms ${\cal L}_{F1}$ and ${\cal L}_{F2}$ we use
the following identity to decompose the product of three $\Gamma$
matrices

$$
\Gamma^a \Gamma^b \Gamma^c = \eta^{ab}\Gamma^c - \eta^{ac}\Gamma^b
+ \eta^{bc} \Gamma^a + i \epsilon^{abcd}\Gamma^{{\dot 5}} \Gamma_d
\eqno{(A.6.2)} \label{A.6.2}
$$

from where we obtain the following simplification using the
symmetric properties of the tensors $\Omega_{BCA}$

$$
\Gamma^a\Gamma^b\Gamma^c \Omega_{bca} ~~=~~  2 \eta^{ab} \Gamma^c
\Omega_{bca} + i \epsilon^{abcd} \Gamma^{{\dot 5}} \Gamma_d
\Omega_{bca}. \eqno{(A.6.3)} \label{A.6.3}
$$

   The second term in Eqn ( A.6.3) yields a vanishing contribution
   when inserted between spinors ( Eqn ( A.6.1) ). Hence we can
   omit it and obtain,
$$
\begin{array}{ccc}
\Gamma^a A_1 &~=~& \Gamma^a E^\mu_a (D_\mu + i B_\mu) + {i \over
4}
\Gamma^a \Gamma^b \Gamma^c \Omega^{(0)}_{bca} \\
&~=~& \Gamma^a [E^\mu_a (D_\mu + i B_\mu) - {i \over 2}\eta^{bc}
\Omega^{(0)}_{abc}]
\end{array}
\eqno{(A.6.4)} \label{A.6.4}
$$

$$
\begin{array}{ccc}
(A_1)_+ &~=~&  e^\mu_a (\nabla_\mu + ib_{\mu+}) - iv_a^\mu
b_{\mu-} - {i \over 2} ( v_a^\mu v_b^\nu - v_a^\nu v_b^\mu)
\partial_\mu e^b_\nu  \\
&~~~& + {i \over 2} ( X^{[\mu,\nu]}_{ab} \partial_\mu (
v^\lambda_\nu s^b_\lambda)
+ Y^{[\mu, \nu]}_{ab} \partial_\mu s^b_\nu), \\
(A_1)_- &~=~&   v^\mu_a (i\partial_\mu - b_{\mu +} ) - e^\mu_a b_{\mu-} \\
&~~~& - {1 \over 2} ( X^{[\mu,\nu]}_{ab}
\partial_\mu s^b_\nu + Y^{[\mu, \nu]}_{ab} \partial_\mu ( e^b_\nu + v^\lambda_\nu s^b_\lambda)),
\end{array}
\eqno{(A.6.5)} \label{A.6.5}
$$
where the general covariant derivative $\nabla_\mu = \partial_\mu
+ {i \over 2} e^\nu_b(\partial_\mu e^b_\mu - \partial_\nu
e^b_\mu)$.

$$
\Gamma^a A_2 =  {i\over 4} \Gamma^a \Gamma^b \Gamma^c
\Omega^{(1)}_{bca} =  {i \over 2} \eta^{ab} \Gamma^c
\Omega^{(1)}_{bca} \eqno{(A.6.6)} \label{A.6.6}
$$

$$
\begin{array}{ccc}
(A_2)_+ &~=~&  i m s^b_\mu[  X^{[\mu,
\nu]}_{ab}a_{\nu-} +  Y^{[\mu, \nu]}_{ab} a_{\nu+} ], \\
(A_2)_- &~=~&  m s^b_\mu [  X^{[\mu, \nu]}_{ab} a_{\nu+} -
Y^{[\mu, \nu]}_{ab} a_{\nu-}],
\end{array}
\eqno {(A.6.7)} \label{A.6.7}
$$

$$
\Gamma^a A_3 = - \Gamma^a E^\mu_a A_\mu (D_5+ i B_5),
\eqno{(A.6.8)} \label{A.6.8}
$$

$$
\begin{array}{ccc}
(A_3)_+ &=& -{ i \over 2 } [ ( e^\mu_a - i v^\mu_a) \eta ( a_{\mu
+} + i a_{\mu -}) + ( e^\mu_a + i v^\mu_a) \bar \eta ( a_{\mu +} -
i a_{\mu -})]\\
(A_3)_- &=& { 1 \over 2 } ( v^\mu_a - i e^\mu_a) \eta (a_{\mu +} +
i a_{\mu -}) + { 1 \over 2 } ( v^\mu_a + i e^\mu_a) \bar \eta
(a_{\mu +} - i a_{\mu -})
\end{array}
 \eqno{(A.6.9)}
 \label{A.6.9}
$$

$$
\Gamma^{{\dot 5}}A_4 = {1 \over 2} \Gamma^{{\dot 5}} [ i (
\Gamma^a\Gamma^b \Omega_{b{\dot 5}a}) + 2  \Phi^{-1}( D_5 + i
B_5)], \eqno{(A.6.10)} \label{A.6.10}
$$

Since the tensor $s^{\mu \nu}$ is symmetric, the terms, which
contain $\gamma^a \gamma^b$, reduce to the trace of that tensor
giving

$$
\begin{array}{ccc}
(A_4)_+ &=& -i(\phi^2_+ + \phi^2_- )^{-1} ( {1 \over 2} \eta (
\phi_+ + i\phi_-) + {1 \over 2} \bar \eta (\phi_+ - i\phi_-)
- m  s^\mu_{\mu} \phi_-)\\
(A_4)_- &=&  (\phi^2_+ + \phi^2_-)^{-1} (  {i \over
2}\eta(\phi_++i\phi_-)+ -{ i\over 2} \bar \eta (\phi_+ - i \phi_-)
- m  s^\mu_{\mu} \phi_+)
\end{array}
 \eqno{(A.6.11)}
 \label{A.6.11}
$$
Here, we give only the final results: \refstepcounter{equation}
$$
\begin{array}{ccc}
{\cal L}_{F1} &~=~& i \bar \Psi (\Gamma^a  E^\mu_a( D_\mu + i
B_\mu) + {i \over 4}\Gamma^a \Gamma^b \Gamma^c
\Omega^{(0)}_{bca}))\Psi ~~~~~~~~~~~~~~~~~~~~~~~\\
&~=~& {i \over 2} \bar \psi \gamma^a[ e^\mu_a (\nabla_\mu +
ib_{\mu+}) - iv_a^\mu b_{\mu-} + {i \over 2} ( v_a^\mu v_b^\nu -
v_a^\nu v_b^\mu)
\partial_\mu e^b_\nu  \\
&~~~& - {i \over 2} ( X^{[\mu,\nu]}_{ab} \partial_\mu (
v^\lambda_\nu s^b_\lambda) + Y^{[\mu, \nu]}_{ab} \partial_\mu
s^b_\nu)] \psi \\
&~& + {i \over 2} \bar \psi \gamma^a \gamma^5[v^\mu_a (i\partial_\mu - b_{\mu+} ) - e^\mu_a b_{\mu-} \\
&~~~& + {1 \over 2} ( X^{[\mu,\nu]}_{ab}
\partial_\mu s^b_\nu + Y^{[\mu, \nu]}_{ab} \partial_\mu ( e^b_\nu + v^\lambda_\nu s^b_\lambda))]
\psi + h.c.
\end{array}
\eqno{(A.6.12)} \label{A.6.12}
$$
\refstepcounter{equation}
$$
\begin{array}{ccc}
{\cal L}_{F2} &~=~&  {i \over 4} \bar \Psi \Gamma^a \Gamma^b
\Gamma^c \Omega^{(1)}_{bca} \Psi~~~~~~~~~~~~~~~~~~~~~~~~~~~~~~~~~~~~~~~~~~~~~~~~~~~~\\
&~=~& {m \over 2} \bar \psi \gamma^a(X^{[\mu,\nu]}_{ab} + i
\gamma^5 Y_{ab}^{[\mu, \nu]})( i\gamma^5 a_{\nu +} -
a_{\nu-})s^b_\mu \psi + h.c.
\end{array}
\eqno{(A.6.13)} \label{A.6.13}
$$
\refstepcounter{equation}
$$
\begin{array}{ccc}
{\cal L}_{F3} &~=~& i \bar \Psi \Gamma^a E_a^\mu A_\mu ( D_5 +
i B_5)\Psi ~~~~~~~~~~~~~~~~~~~~~~~~~~~~~~~~~~~~~\\
&~=~& - { 1 \over 2} \bar \psi \gamma^a \eta( a_{\mu+} + i
a_{\mu-} ) [ ( e^\mu_a -i v^\mu_a) + \gamma^5 ( e^\mu_a +
iv^\mu_a)] \psi + h.c.
\end{array}
\eqno{(A.6.14)} \label{A.6.14}
$$
\refstepcounter{equation}
$$
\begin{array}{ccc}
{\cal L}_{F4} &=&  {i \over 2} \bar  \Psi \Gamma^{{\dot 5}} [ (
\Gamma^a \Gamma^b \Omega_{b{\dot 5}a}) + 2 \Phi^{-1}( D_5 + i
B_5)]
\Psi ~~~~~~~~~~~~~~~~~~~~~~~~~~~~~~~~~~~~\\
&=& - {i \over 2} \bar \psi ( \phi_+^2 + \phi_-^2)^{-1} \eta (
\phi_+ + i\phi_-)(1 - \gamma^5) \psi + h.c.
\end{array}
\eqno{(A.6.15)} \label{A.6.15}
$$

\end{document}